\documentclass[
aip, 
amsmath,amssymb,
preprint,
]{revtex4-1}

\usepackage{graphicx}
\usepackage{dcolumn}
\usepackage{bm}
\usepackage{xcolor}

\usepackage[utf8]{inputenc}
\usepackage[T1]{fontenc}
\usepackage{etoolbox}

\makeatletter
\def\@email#1#2{
 \endgroup
 \patchcmd{\titleblock@produce}
  {\frontmatter@RRAPformat}
  {\frontmatter@RRAPformat{\produce@RRAP{*#1\href{mailto:#2}{#2}}}\frontmatter@RRAPformat}
  {}{}
}
\makeatother
\usepackage{mathrsfs}

\newcommand{\elec}{\text{e}}
\newcommand{\gas}{\text{g}}
\newcommand{\ion}{\text{i}}
\newcommand{\el}{\text{el}}
\newcommand{\iz}{\text{iz}}
\newcommand{\nC}{c_\star}
\newcommand{\partialx}[1][]{\frac{\partial{#1}}{\partial x}}
\newcommand{\vth}{v_\text{th}}
\newcommand{\change}[1]{#1}

\begin{document}

\title[Comparison of high-order moment models for ions]{Comparison of high-order moment models for the ion dynamics in a bounded low-temperature plasma}

\author{A. Berger}
\email{anatole.berger@lpp.polytechnique.fr}
\affiliation{Laboratoire de Physique des Plasmas, Centre National de la Recherche Scientifique, Sorbonne Université, École polytechnique, Institut Polytechnique de Paris, route de Saclay 91128 Palaiseau, France}
\affiliation{Aeronautics and aerospace department, von Karman Institute for Fluid Dynamics, Waterloosesteenweg 72,
B-1640 Sint-Genesius-Rode, Belgium}
\author{\change{N. Lequette}}
\affiliation{Laboratoire de Physique des Plasmas, Centre National de la Recherche Scientifique, Sorbonne Université, École polytechnique, Institut Polytechnique de Paris, route de Saclay 91128 Palaiseau, France}
\author{T. Magin}
\affiliation{Aeronautics and aerospace department, von Karman Institute for Fluid Dynamics, Waterloosesteenweg 72,
B-1640 Sint-Genesius-Rode, Belgium}
\author{A. Bourdon}
\affiliation{Laboratoire de Physique des Plasmas, Centre National de la Recherche Scientifique, Sorbonne Université, École polytechnique, Institut Polytechnique de Paris, route de Saclay 91128 Palaiseau, France}
\author{A. Alvarez Laguna}
\affiliation{Laboratoire de Physique des Plasmas, Centre National de la Recherche Scientifique, Sorbonne Université, École polytechnique, Institut Polytechnique de Paris, route de Saclay 91128 Palaiseau, France}

\begin{abstract}
{\small
Low-temperature plasmas often present non-equilibrium ion distribution functions due to the collisions with the background gas and the presence of strong electric fields. This non-equilibrium is beyond classical fluid models, often requiring computationally-intensive kinetic simulations. In our work, we study high-order moment models in order to capture the non-equilibrium state with a macroscopic set of equations, which is more computationally efficient than kinetic simulations. We compare numerical simulations of different moment closures: Grad's closure, the hyperbolic quadrature method of moments (HyQMOM), the extended quadrature method of moments, and a method based on entropy maximization. We assess the different closures for plasma applications and propose efficient numerical discretizations. The numerical solution of the high-order moment models is compared to kinetic simulations of an argon plasma between two floating walls at different pressure regimes, from nearly collisionless to collisionally-dominated. In general, all the high-order moment closures capture the ion transport with high fidelity as compared to the kinetic simulations, providing an improvement as compared to classical fluid models. Classical fluid closures such as the Fourier law for the heat flux is shown to be not suitable to capture the sheath or the low pressure regime. In addition, the ability of each moment method to reconstruct the velocity distribution function from the moments is assessed. The high-order moment models are able to capture the non-equilibrium distributions in the bulk and sheath with remarkable fidelity, dramatically improving classical fluid models while having comparable computational cost. In particular, the HyQMOM shows to be a robust method that provides an excellent comparison with the kinetic simulations of both the moments and the distribution function in the bulk and the sheath.}
\end{abstract}

\maketitle

\section{Introduction}

Low-temperature plasmas are used in a wide variety of applications such as electric propulsion \cite{Goebel08}, plasma etching for microelectronics \cite{Liberman05,Chabert11}, plasma-assisted combustion \cite{Starikovskaia06}, and biomedical applications\cite{Kong09}. In general, these plasmas present a low ionization level that results in conditions far from thermodynamical equilibrium, where the charged species collide more often with the neutral species of the gas than among themselves. In addition, low-temperature plasmas are often in contact with the walls of the reactor and other surfaces. The presence of the walls creates a plasma sheath with strong gradients and electric fields, driving the plasma species even further from thermodynamic equilibrium. For these reasons, the charged species present strongly non-Maxwellian distribution functions in the sheath regions, which plays an important role for some applications such as plasma etching or surface treatment, where the ion energy distribution has a large impact in the surface processes~\cite{Kawamura99}. 

Classical fluid models solve the particle number, momentum, and energy balance equations, i.e., 3 moments. The validity of classical fluid models for modeling these type of non-equilibrium plasmas is limited since, strictly speaking, they only consider Maxwellian velocity distribution functions (VDF) or small perturbations from the thermodynamic equilibrium, and are therefore unable to capture phenomena such as strong heat fluxes, pressure anisotropies or the non-Maxwellian VDF in the sheath. Despite these limitations, different theoretical models based on the fluid formalism have been proposed\cite{Schottky24,Godyak86} to describe bounded low-temperature plasmas (a summary can be found in Refs.\cite{Chabert11, Liberman05}). More recent works\cite{Tavant19, Beving22}, have revisited these fluid models based on kinetic simulations. The kinetic \change{simulations (such as particle-in-cell simulations\cite{Vahedi95,Birdsall91}, semi-Lagrangian methods\cite{Sonnendrucker99}, Boltzmann solvers\cite{Gamba18,Oblapenko20} or Fokker-Plank solvers\cite{Kolobov03})} are valid in the whole range of pressures but computationally much more expensive than fluid models. However, most of the fluid models used in the theory of bounded plasmas still consider a reduced number of moments (e.g., often assuming isothermal conditions or at most following empirical polytropic laws \cite{Kuhn10, Tavant19}). In addition, since they are based on a Maxwellian VDF, they are not able to \change{provide a} better approximation of the non-equilibrium distribution function of the ions at the wall.

A potential alternative to kinetic models is the method of moments. This method extends the fluid variables by considering higher-order moment equations, e.g., pressure tensor, heat-flux, etc. As a result the method of moments can capture more general VDFs than classical fluid models, i.e. non-Maxwellian distributions, hence extending the validity of the model to non-equilibrium conditions. In addition, moment methods are, in general, computationally more efficient than kinetic models and without the statistical noise of particle-based kinetic methods. The difficulty associated to these models is in providing a consistent, mathematically stable, and numerically efficient model for the closing flux that appears in the last equation of the moment hierarchy as well as for the reconstruction of the VDF from the moments. Various closures for the moment equations have been proposed for different applications (see e.g. Ref.\cite{Pichard_review} for a recent review). In particular, in partially ionized plasmas, Zhdanov~\cite{Zhdanov03} used Grad's moment method to derive general transport coefficients of neutral and charged species, in the linearized moment regime (i.e., close to thermodynamic equilibrium), also recently revisited by Hunana\cite{Hunana25}. Under conditions further from equilibrium we find some recent numerical works. Bocelli et al.~\cite{Boccelli20,Boccelli22} have applied an interpolative maximum-entropy closure to electrons and ions in a magnetized low-temperature plasma. Taunay et al.~\cite{Taunay23} considered different quadrature methods of moments (QMOM) for a collisionless Vlasov-Poisson system. Alvarez Laguna et al.~\cite{Alvarez22,AlvarezLaguna23} developed a regularized Grad's model for electrons in a reacting partially-ionized plasma, later generalized in Ref.\cite{AlvarezLaguna25}. Kuldinow et al.~\cite{Kuldinow24} have proposed a ten-moment model with a heat-flux closure. However, despite these efforts, the realistic modeling of the multicomponent collisional processes in the moment equations as well as the behavior of these models at different pressures in the presence of sheaths is still an open question for the moment closures. Similarly, the comparison with kinetic simulations of a bounded low-temperature plasma, representing the plasma-sheath transition, is still a challenge due to the presence of strong electric fields in the plasma sheath~\cite{Alvarez20,Hara20}. In addition, as mentioned above, the ability of moment models to capture the ion VDF inside the sheath has not been studied so far.

In this paper, we focus on the ion dynamics in a one-dimensional argon plasma between two floating walls in order to represent the plasma-sheath transition. This work follows a similar methodology to the work of Laplante and Groth\cite{Laplante16}, where different high-order moment \change{models} were compared for the representation of a shock in a monoatomic rarefied gas. However, in low-temperature plasmas, the sources of non-equilibrium and the numerical challenges for the moment models are different to a shock in monoatomic rarefied gases. In the shock test case of Laplante and Groth, the non-equilibrium is due to a gradient that is smaller than the characteristic particle-like collisional mean-free path, which leads to smooth shock jumps. Furthermore, the hypersonic velocities are a challenge for the moment models in order to correctly capture the smooth jump. Alternatively, in low-temperature plasmas, due to the weakly-ionized state, the collisions between charged particles are often negligible as compared to the collisions of the charged particles with the background gas. As a result, even in collisionally-dominated plasmas at high pressures, the ions can be out of thermodynamic equilibrium when the plasma is weakly ionized. Another source of non-equilibrium \change{are} the \change{charge-exchange} collisions, which is often dominant in the ion collisions, and introduces an important anisotropy in the collision geometry, not present in the collision model of Laplante and Groth. Finally, the electric field plays a fundamental role in the non-equilibrium of low-temperature plasmas. In particular, it is very strong in the sheath region, and leads to non-Maxwellian distributions with distributions that strongly depends on the pressure regime. These challenges are very specific to partially-ionized plasmas and will be studied in this paper.

In our paper, we will compare different high-order moment closures, some of them never studied for low-temperature plasmas. We will consider: a regularized Grad's moment closure\cite{Cai14}, an interpolative maximum entropy closure\cite{McDonald13,Baradaran15, Boccelli24}, and two QMOM models: the hyperbolic QMOM (HyQMOM)\cite{Fox18} and the extended QMOM (EQMOM)\cite{Chalons10}. For each method, we will solve up to the fourth-order moment, i.e., five scalar moments (5M) in 1D. As some of the closures do not allow for an analytical integration of the collisional processes, we will study a simplified collisional model, considering only ionization and charge-exchange collisions. We will propose an efficient numerical resolution of the moment equations that is based on analytical formulas or approximations of the information required by the numerical scheme (such as the eigenvalues of the flux Jacobian or the closing flux), minimizing the computational cost, which remains comparable to classical 3M models.

The paper is organized as follows. In Section~\ref{sec:MOM}, we will introduce the setup representing a low-pressure plasma plasma between two floating walls. Then, we will present the underlying kinetic model for the ions. After, we will present the moment equations as well as the four considered high-order moment closures, focusing on the theoretical aspects and requirements for a robust moment set of equations. In Section~\ref{sec:num}, we will present the numerical scheme for the different set of equations. In particular, we will \change{provide} analytical and approximated expressions for the numerical fluxes, which allows to have very efficient numerical solvers. Finally, in Section~\ref{sec:results}, we will present the numerical results of the different moment closures representing the ion dynamics in an argon plasma between two floating walls. We will compare the solutions of the moment models to kinetic simulations, comparing both the representation of the moments and the reconstructed VDF in the bulk and the sheath. The paper will finish with a discussion of the findings and a summary of the conclusions in Section \ref{sec:conclu}.

\section{Kinetic and moment models \label{sec:MOM}}

\subsection{Numerical setup and kinetic simulations \label{sec:NumericalSetUp}}

\begin{figure}[htbp]
    \centering
    \includegraphics[width=0.5\linewidth]{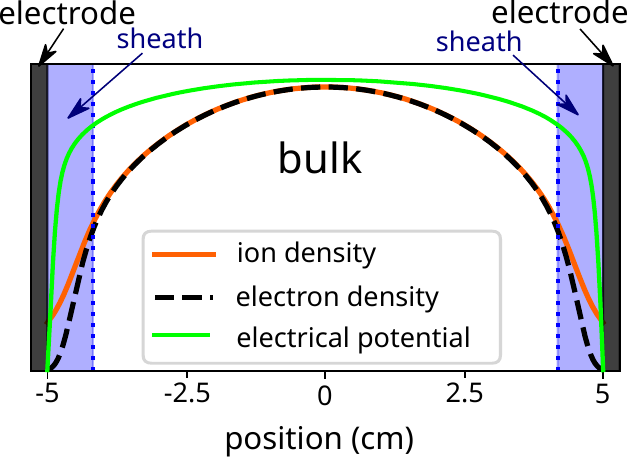}
    \caption{Scheme of the numerical set-up.}
    \label{fig:scheme_simu}
\end{figure}

We consider a one-dimensional partially-ionized plasma between two floating walls. The numerical set-up mimics an electropositive noble gas discharge at different pressure regimes \cite{Chabert11, Liberman05}. A scheme of the setup is presented in Fig.~\ref{fig:scheme_simu}. We show the ion and electron densities (in black and orange, respectively), along with the electrical potential (in green). 
\change{
The system exhibit two regions 
with very different dynamics:} 
the bulk, in the center of the discharge, where the potential is mostly flat and densities of ions and electrons are nearly equal; and the sheath, in the vicinity of the wall, where the plasma is non-neutral (positively charged), leading to a sharp drop of the electrical potential that will accelerate the ions to very high velocity towards the wall. 

The kinetic simulations that will be used as a reference for the high-order moment model are based on the particle-in-cell Monte-Carlo collisions (PIC-MCC) method. We study a domain that extends from $x\in[-L,L]$, where the length $L=5$~cm. The electrons and singly-ionized ions are simulated with the PIC-MCC method whereas the neutral gas is considered to be a spatially-homogeneous background at different pressures with constant temperature of $T_\gas=300$ K. The plasma species are initialized with a constant profile at densities $n_\elec(t=0)=n_\ion(t=0)=10^{15}$ m$^{-3}$ and $T_\elec = 5$ eV and $T_\ion=T_\gas$, where the subscripts $\elec,\,\ion$ and $\gas$ refer to electrons, ions, and gas, respectively. The charged species are absorbed at the boundaries of the simulation domain, where the electric potential is imposed to be $\phi(x=-L)=\phi(x=L)=0$. In order to balance the loss of charged species at the walls, a number of electrons and ions that is equal to the number of ions lost at the walls is injected at every time step with a probability that is proportional to the electron local density, as explained in Ref.~\cite{Alvarez20}. The electrons are injected at $T_{inj}$ (given in Table \ref{tab:NumParams}) and the ions at the neutral temperature $T_\gas$, both species injected following a Maxwellian distribution. This particle injection method mimics both the ionization process for the creation of ions and the electron heating mechanism that sustains the discharge. In this paper, we will focus on the ion kinetics and the considered injection model is effectively equivalent to an electron-impact ionization for the ion population. Similar one-dimensional set-ups have been previously considered by different authors to study the plasma-sheath transition in gas discharges \cite{Tavant19, Alvarez20, Beving22, Gangemi25}.

In our study, we will compare different high-order moment closures for ions to the kinetic simulations. Some of the moment closures do not allow for analytical integration of the collisional terms. For this reason we will consider a simplified model for the collisional cross sections in an argon plasma. The ion collisions with the neutrals will be modeled with a back-scattering collision with a polarization potential model where the cross section is proportional to the inverse of the relative velocity, as described in the next section. Coulomb collisions are neglected in the model. The PIC code used in this work is 1D-3V but, as it will be shown in the following section, the ion kinetic equation of the above-mentioned set-up can be reduced to a 1D-1V problem. The electrons carry out elastic collisions with the gas (ionization and excitation collisions are not included in the model), with the elastic cross section as found in the Phelps database from LXCat \cite{PhelpsLX}. 

As explained by classical works on the plasma-sheath formation, based on the isothermal fluid equations with Boltzmann distributed electrons\cite{Riemann92, Riemann05, Chabert11}, the studied plasma equilibrium depends mainly on the following non-dimensional parameters: The ratio between the Debye length and the domain length $\lambda_\text{D}/L$, with $\lambda_D = \sqrt{\varepsilon_0 T_\elec/(n_\elec e)}$ where $\varepsilon_0$ is the vacuum permittivity, $T_\elec$ is the electron temperature, $n_\elec$ is the electron density, and $e$ is the elementary charge; the ratio between the ion mean free path and the domain length $\lambda_\text{mpf}/L$, with $\lambda_\text{mpf} = (n_\gas\sigma_0 )^{-1}$ where $n_\gas$ is the gas density, $\sigma_0 = 10^{-18}$ m$^{2}$ is a reference ion-neutral collision cross section for argon\cite{Chabert11}; and the ratio between the ion and electron temperatures $T_\ion/T_\elec$. A summary of the numerical parameters as well as the above-mentioned normalized parameters for the different cases of study are summarized in Table \ref{tab:NumParams}. 

\begin{table}
    \begin{tabular}{c | c c c c | c c c }
        pressure (Pa) & $N_\text{ppc}$ & $N_{x}$ & $T_\text{g}$ (eV) & $T_\text{inj}$ (eV) &$\lambda_\text{mpf}/L$ &  $\lambda_\text{D}/L$ & $T_\text{i}/T_\text{e}$\\ [0.5ex] 
        \hline\hline
        0.01 & $100$ & $2000$ & 0.025 & 5 & 10 & $2\cdot 10^{-3}$ & 0.04 \\ 
        0.1  & $100$ & $2000$ & 0.025 & 5 & 1 & $2\cdot 10^{-3}$ & 0.06\\ 
        1    & $100$ & $2000$ & 0.025 & 5 & 0.1 &  $2\cdot 10^{-3}$ & 0.05 \\ 
        10   & $200$ & $2000$ & 0.025 & 10 & 0.01 &  $2\cdot 10^{-3}$ & 0.07 \\ [1ex]
        \hline
    \end{tabular}
    \caption{Parameters of the simulations. $N_\text{ppc}$ is the number of particle per cell, $N_x$ the number of cells, $T_\text{g}$ the gas temperature (considered constant and uniform during the simulation), $T_\text{inj}$ the injection temperature of the electrons, $\lambda_\text{mpf}$ the mean-free-path of the ions, $L$ the characteristic length of the system (corresponding to the length of half of the domain), $\lambda_\text{D}$ the Debye length, and $T_\text{i}$ and $T_\text{e}$ the ion and electron temperatures.}
    \label{tab:NumParams}
\end{table}

\subsection{Ion kinetic equation}

As mentioned previously, we focus on the ion kinetics. To simplify notation, we will remove the ion index in the rest of the paper. The ion kinetic model is based on the Boltzmann equation, which describes the evolution in time and space of a VDF $f(t, \vec{r}, \vec{v})$ of a particle of mass $m$ and charge $e$, under an external electric field $\vec{E}$, as follows,
\begin{eqnarray}\label{eq:kinetic}
    \frac{\partial{f}}{\partial t} + \vec{v}\cdot \vec{\nabla}_r f + \frac{e\vec{E}}{m}\cdot \vec{\nabla}_v f &=& \left. \frac{\delta f}{\delta t}\right\rvert_\text{c} = \left. \frac{\delta f}{\delta t}\right\rvert^\el_\text{c} + \left. \frac{\delta f}{\delta t}\right\rvert^\iz_\text{c},
\end{eqnarray}
where the right-hand side term account for the collisional processes, divided into ion-gas elastic collisions and ionization collisions (the self collisions and recombination \change{collisions} are neglected as the plasma is considered to be weakly-ionized and therefore these collisions are negligible as compared to the ion-gas collisions). The ion-gas elastic collisions \change{are} modeled with the Boltzmann collision term:
\begin{eqnarray}
    \left. \frac{\delta f}{\delta t}\right\rvert^{\el}_\text{c} &=& \int\int \left[f(\vec{v}') f_\text{g}(\vec{v}'_\text{g}) - f(\vec{v}) f_\text{g}(\vec{v}_\text{g})\right]\sigma(\lvert \vec{v}-\vec{v}_\text{g}\rvert, \chi) \lvert \vec{v}-\vec{v}_\text{g}\rvert \,\text{d}^2\Omega \text{d}^3v_\text{g},
\end{eqnarray}
where the primed velocities refer to the velocities of the restitution collision (which conserves momentum, energy and follows micro-reversibility of the collision), $\sigma$ is the differential cross section, $\chi$ is the scattering angle, and $\text{d}^2\Omega$ is the unit sphere element of the collision angles.

As explained by Robson et al.\cite{Robson17}, 
\change{the charge-exchange collision can be approximated as purely backscattering, i.e., the scattering angle is $\chi\simeq \pi$, and the differential cross section will have the following expression:} 
\begin{equation}
    \change{\sigma(\lvert \vec{v}-\vec{v}_\text{g}\rvert, \chi)=\frac{\sigma^{(0)}(\lvert \vec{v}-\vec{v}_\text{g}\rvert)}{2\pi}\delta(\cos\chi-1),}
\end{equation}
where $\delta$ is the Dirac delta function and $\sigma^{(0)}(\lvert \vec{v}-\vec{v}_\text{g}\rvert)$ is the total cross section.

If we consider a Langevin potential, i.e., a polarization potential, the cross section is proportional to the inverse of the relative speed, and the integrals over the scattering angle and the gas velocity can be performed, as follows,
\begin{eqnarray}\label{eq:ElCollIon}
    \left. \frac{\delta f}{\delta t}\right\rvert^{\el}_\text{c} &=& K^{(0)} \int \left[f(\vec{v}_g) f_\text{g}(\vec{v}) - f(\vec{v}) f_\text{g}(\vec{v}_\text{g})\right] \,\text{d}^3v_\text{g} = n_\gas K^{(0)}\left[n(\vec{r},t)\,w_\gas(\vec{v}) - f(\vec{v})\right] \, ,
\end{eqnarray}
where $K^{(0)}$ is a collision rate, such that $\sigma^{(0)}(\lvert \vec{v} - \vec{v}_\text{g}\rvert) = \frac{K^{(0)}}{\lvert v - v_\text{g}\rvert}$. In this work, we choose $K^{(0)} = 4.6\cdot 10^{-16}$ m$^3\cdot$s$^{-1}$ so that the simplified cross section has a similar magnitude to the actual argon cross section. In Eq.~\eqref{eq:ElCollIon}, $n = \int f(\vec{v})\text{d}^3v$ is the ion density, and $w_\gas(\vec{v}) = f_\gas(\vec{v})/n_\gas$. As seen in Eq.~\eqref{eq:ElCollIon}, the resulting collision operator is equivalent to a Bhatnagar–Gross–Krook (BGK) operator. 

The ionization collision is expressed as follows:
\begin{equation}
    \left. \frac{\delta f}{\delta t}\right\rvert^\iz_\text{c} = n_\gas K^{(0)}_\iz\, n_\elec(\vec{r},t)\,w_\gas(\vec{v}),
\end{equation}
where the ionization rate is such that, at each time step, the number of particles added to the domain is equal to the number of particles that left the domain at the boundaries. 
In other words,
\begin{equation}
    \change{
        K_\text{iz}^{(0)} = \frac{\Gamma_{\text{i,wall}}(x=-L) + \Gamma_{\text{i,wall}}(x=L)}{n_\text{g} \int_{-L}^L n_\text{e}(x) \,\text{d}x},
    }
\end{equation}
where $\Gamma_\text{i,wall}$ is the absolute value of the ion flux at each wall. 

In the resulting kinetic model, due to the angular geometry of the back-scattering collision and the 1D geometry, the velocities of the ions in the perpendicular direction are decorrelated and hence the kinetic model can be reduced to a 1D-1V problem, as the distribution function is in equilibrium at the gas temperature in the transverse direction.  In the following, to simplify notation, we will denote $\vec{v} = v\vec{e}_x+v_\perp\vec{e}_\perp$. 
In that same spirit, we define the one-dimensional VDF that depends on the velocity along $x$, $f(v)$, defined as
\begin{equation}
    f(v) = \int_{\mathbb{R}^2} f^\text{3D}(\vec{v}) \,\text{d}v_y \,\text{d}v_z \,.
\end{equation}

\subsection{One-dimensional high-order moment equations }

The moment equations are obtained by taking averages over the velocity space of the kinetic equation, i.e., Eq.~\eqref{eq:kinetic}. 
In the high-order moment theory, it is useful to use different definitions for the moments, as it helps for the notation and analysis of the results. First, we define the velocity moments of order $n$ as,
\begin{eqnarray}\label{eq:momentCons}
    M^{(n)} &=& \int_{-\infty}^\infty m v^n f(v) \,\text{d}v \,.
\end{eqnarray}
The first moments are: $M^{(0)} = \rho = mn$ the mass density, and $M^{(1)} = \rho u$ the mass density flux with $u$ the drift velocity. 
\change{Secondly}, we define the centered moments \change{as}
\begin{eqnarray}
    P^{(n)} &=& \int_{-\infty}^\infty m c^n f(c) \,\text{d}c \quad \text{with} \quad c = v-u.
\end{eqnarray}
Finally, in the moment theory it is useful to define the normalized (or standarized) moments as follows:
\begin{eqnarray}
    P_\star^{(n)} &=& \frac{m \change{\vth}}{\rho}\int_{-\infty}^\infty \nC^n f(\nC) \,\text{d}\nC = \frac{P^{(n)}}{\rho \change{\vth}^n} \quad \text{with} \quad \nC = \frac{v-u}{\change{\vth}},
\end{eqnarray}
where the thermal velocity is defined as $\change{\vth} = \sqrt{p/\rho}$, with $p = P^{(2)}$ \change{being} the pressure.

In our study, we will work with 5M closures, which solve the equations up to the fourth-order moment (for $n=\{0,\cdots,4\}$ in Eq.\eqref{eq:momentCons}) and the fifth-order moment is the closing flux, needed to truncate the moment hierarchy. 
\change{Indeed, we found that the fifth-order moment as seen in PIC-MCC simulations depends on the third and fourth order moment in a manner that was independent of the pressure\cite{Berger25RGD}. 
Hence, there was a possibility to express the fifth-order moment as a function of the previous moments, which motivates one to consider it as the closing moment.} 

We define the 5M centered moments as:
\begin{equation}
    P^{(2)} \equiv p \,, \quad\quad P^{(3)} \equiv q \,, \quad\quad  P^{(4)} \equiv r \quad\quad \text{and} \quad P^{(5)} \equiv s,
\end{equation}
respectively, the pressure, the heat flux \change{(or skewness)}, the kurtosis, and the \change{hyperskewness}\cite{Hunana22}. Note that as we are considering a 1D-1V problem, the pressure represents the $xx$ component of the pressure tensor, the heat-flux the $xxx$ component of the third-rank heat-flux tensor and so forth.

The standardized first six moments are denoted as:
\begin{eqnarray}
    P_\star^{(0)} = 1 \,,\quad\quad P_\star^{(1)} = 0 \,,\quad\quad P_\star^{(2)} = 1 \,,\quad\quad P_\star^{(3)} \equiv q_\star \,,\quad\quad P_\star^{(4)} \equiv r_\star \,,\quad\quad P_\star^{(5)} \equiv s_\star.
\end{eqnarray}
Finally, the velocity moments are related to the centered moments by the following relations,
\begin{subequations}\label{eq:consMoments5M}
\begin{eqnarray}
M^{(2)} &=& \rho u^2 + p,  \\
M^{(3)} &=&\rho u^3 + 3 u p + q,  \\
M^{(4)} &=&\rho u^4 + 6 u^2 p + 4 u q + r, \\
M^{(5)} &=&\rho u^5 + 10 u^3 p + 10 u^2 q + 5ru + s.
\end{eqnarray}
\end{subequations}

The moment equations are obtained by taking moments of the kinetic equation, Eq.~\eqref{eq:kinetic}, which \change{yields the} following non-linear set of equations,
\begin{subequations}\label{eq:system5M}
\begin{eqnarray}
    \frac{\partial \rho }{\partial t}+ \partialx[(\rho u)] &=& S_\text{iz} \label{eq:5M-0} \\
    \frac{\partial (\rho u)}{\partial t} + \partialx[M^{(2)}] &=&\frac{e E}{m} \rho+ \mathcal{C}^{(1)}  \label{eq:5M-1} \\
    \frac{\partial M^{(2)}}{\partial t} + \partialx[M^{(3)}] &=& 2 \frac{e E}{m} \rho u + \mathcal{C}^{(2)} + S_\text{iz} \frac{k_\text{B} T_\text{g}}{m}   \label{eq:5M-2}\\
    \frac{\partial M^{(3)}}{\partial t} + \partialx[M^{(4)}] &=& 3 \frac{e E}{m} M^{(2)} + \mathcal{C}^{(3)}  \label{eq:5M-3} \\
    \frac{\partial M^{(4)}}{\partial t} + \partialx[M^{(5)}] &=& 4 \frac{e E}{m} M^{(3)} + \mathcal{C}^{(4)} + 3 S_\text{iz} \left(\frac{k_\text{B} T_\text{g}}{m}\right)^2 \label{eq:5M-4}
\end{eqnarray}
\end{subequations}
where $\mathcal{C}^{(n)}$ are the collision terms due to elastic collisions and $S_\text{iz}=mn_gn_eK_\text{iz}^{(0)}$ is the rate of ion density creation by ionization. 

For the elastic collision terms, we take the moments of Eq.~\eqref{eq:ElCollIon}, which yields,
\begin{eqnarray}
    \mathcal{C}^{(1)} = -\frac{K^{(0)}}{m} \rho_\text{i} \rho_\text{g} u_{\text{i}}, &\quad\quad & \mathcal{C}^{(2)} = -\frac{K^{(0)}}{m} \left( \rho_\text{g} M_{\text{i}}^{(2)} - \rho_\text{i} p_\text{g} \right), \nonumber \\
    \mathcal{C}^{(3)} = -\frac{K^{(0)}}{m} \rho_{\text{g}} M_{\text{i}}^{(3)}, &\quad\quad & \mathcal{C}^{(4)} = -\frac{K^{(0)}}{m} \left(\rho_\text{g} M_{\text{i}}^{(4)} - 3 \rho_\text{i} p_\text{g}^2 \right).
\end{eqnarray}

\subsection{Analytical and interpolative 5M closures}

The system of equations \eqref{eq:system5M} is not closed as the moment $M^{(5)}$ (or more precisely the centered moment $s$, also known as hyper skewness) needs to be written as a function of the other "known" moments. This is the so-called closure problem in the moment theory. 
The usual way to do so is to consider a VDF expression that depends on five parameters. In general, these five parameters depend non-linearly on the first five moments and require a numerical inversion of a system of equations. As a result, once \change{the numerical system has been inverted}, the VDF is obtained and used to compute the closing moment. The mathematical expression for the VDF as a function of the moments should respect several properties in order to be used as a moment closure: (i) since the VDF represents a probability density, it should be positive in all the velocity domain (here $\mathbb{R}$) for all sets of parameters in their definition space; (ii) the moment system inversion should have a solution for any set of moments in the realizability space (defined as the space of moments that provide a positive distribution function, which is independent of the closure); and (iii) the existence of an entropy inequality of the closed set of equations is important for the stability of the system of equations and the uniqueness of the solution in discontinuities\cite{Lax73}. \change{From} a numerical point of view, the system should be hyperbolic (which is related to the existence of an entropy\cite{Lax73}), otherwise the system will have complex (non-real) eigenvalues, which can lead to nonphysical behavior.  Note that all these properties are not necessarily satisfied by all the  common moment closures. 

In this study, we consider four of the most common closures for 5M equations, where we provide analytical expressions or interpolative approximations of the closure.

\subsubsection{Globally hyperbolic regularized Grad closure}

The distribution function of Grad's method of moment is a Maxwellian perturbed by a polynomial in the velocity, which can be conveniently expressed as function of the Hermite polynomials\cite{Grad49}, as follows, 
\begin{eqnarray}
    f_\text{Grad}(c) &=& \mathcal{M}(c; \rho, T) \left( 1 + \sum_{k=0}^{N-1} h_k H^{(k)}(c) \right),
\end{eqnarray}
where $\mathcal{M}(c; \rho, T)$ is a centered Maxwellian of density $\rho$ and temperature $T$, $H^{(k)}$ is the Hermite polynomial of order $k$, and $h_k$ is a coefficient that is a linear combinations of the standarized moments. The inversion is therefore analytical and straightforward, due to the orthogonality of the Hermite polynomials\cite{Balescu88}. 
In the 5M case, the Grad distribution function reads
\begin{equation}\label{eq:5MGrad}
    f_\mathrm{Grad}^\text{5M}(\nC) = \frac{\rho}{m \change{\vth}}\mathcal{M}(\nC) \left( 1 - \frac{q_\star}{6} \left(3\nC - \nC^3\right) + \frac{r_\star - 3}{\change{24}} \left(3 - 6\nC^2 + \nC^4\right) \right),
\end{equation}
where the normalized Maxwellian distribution (or Gaussian distribution) is defined as,
\begin{equation}\label{eq:NormalizedMaxwellian}
    \mathcal{M}_\star(\nC) = \frac{e^{-\nC^2/2}}{\sqrt{2\pi}}\,.
\end{equation}
The standardized closing flux can be easily computed as 
\begin{equation}\label{eq:closingGrad}
    s_\star = \frac{m \change{\vth}}{\rho}\int_{-\infty}^\infty \nC^5 f_\mathrm{Grad}^\text{5M}(\nC) d\nC = 10 q_\star \, .
\end{equation}
In Fig.~\ref{fig:realizability_eachMOM}, the values of the closing flux is presented in the realizability space and compared to other closures.
Despite the simplicity of Grad's distribution function, it has major drawbacks in its applicability to conditions far from equilibrium. First, as seen in Eq.~\eqref{eq:5MGrad}, the distribution function is not guaranteed to be non-negative, which can lead to negative populations in the tail of the VDF. In addition, when the closing flux of Eq.~\eqref{eq:closingGrad} is used in the system of equations \eqref{eq:system5M}, the resulting system has mathematical problems related to the hyperbolicity of the equations. In this paper, we will \change{use} a fix to this problem, proposed by Cai et al.~\cite{Cai14}, which consists in adding an additional term to Eq.~\eqref{eq:5M-4}, allowing to recover a hyperbolic system of equations. As a result, in the globally hyperbolic regularized Grad \change{model}, Eq.~\eqref{eq:5M-4} becomes,
\begin{equation}
    \frac{\partial M^{(4)}}{\partial t} + \partialx[M^{(5)}] + 10\frac{pq}{\rho^2}\partialx[\rho]-5\mathcal{K}\partialx[u] - 10\frac{q}{\rho}\partialx[p]= 4 \frac{e E}{m} M^{(3)} + \mathcal{C}^{(4)} + 3 S_\text{iz} \left(\frac{k_\text{B} T_\text{g}}{m}\right)^2,
\end{equation}
where $\mathcal{K} = r - 3p^2/\rho$. 
\change{Note that the regularization does not change the closure (that remains the regular closure of Grad). This is simply a fix for solving the global system (Eq.~\eqref{eq:system5M}) without loss of hyperbolicity.}

\subsubsection{Interpolative Maximum Entropy}

The maximum entropy closure \cite{Dreyer87, Levermore96} is based on the thermodynamical principle of entropy maximization. We underline that the maximum-entropy principle differs in multi-component mixtures as compared to monoatomic non-reacting gases, as explained in Ref.~\cite{Milana13}. 
Nevertheless, in this work, we take the simple monoatomic gas maximum-entropy principle, as done previously~\cite{Boccelli20,Boccelli22}. 
For a given number of moments, the VDF that maximizes the entropy is an exponential of a polynomial in the velocity. 
In our 5M case, the VDF will thus take the form
\begin{eqnarray}
    f_{ME}^\text{1D}(\nC) &=& \exp\left(\sum_{i=0}^4 k_i \nC^i\right).
\end{eqnarray}
The moment inversion of this closure has no analytical expression and its numerical resolution is stiff and, in general, badly conditioned when far from equilibrium. 
In the results (Sec.~\ref{sec:results}), we will show a comparison between the VDF as reconstructed by the maximum entropy closure and the PIC simulations. 
In order to compute the maximum entropy distribution from the moments, we solve an optimization problem, as explained in Appendix \ref{sec:MEInversion}.

For the computation of the closing flux, however, McDonald and Torrilhon\cite{McDonald13} developed an interpolative formula that approximates the exact inversion with very high accuracy. 
This interpolative formula can be written as
\begin{eqnarray}\label{eq:sME}
    s_\star &=& \frac{q_\star^3}{\beta^2} + \left(10 - 8 \sqrt{\beta}\right) q_\star,
\end{eqnarray}
with
\begin{eqnarray}
    \beta &=& \frac{1}{4} \left(3 - r_\star + \sqrt{(3-r_\star)^2 + 8 q_\star^2}\right) \, .
\end{eqnarray}
The variation of $s_\star$ as a function of $q_\star$ and $r_\star$ is shown in Fig.~\ref{fig:realizability_eachMOM} along with the other closures. 

This closure has the disadvantage of presenting a singular subspace in the realizability space\cite{Junk98}, referred to as the Junk line. 
Indeed, for $ r_\star>3$, if $q_\star = 0$ (which corresponds to the vertical semi-line above the equilibrium in Fig.~\ref{fig:realizability_eachMOM}), the closing moment and the VDF are not defined. Furthermore, on each side of this singular line, the closing \change{hyperskewness} has opposite signs and are diverging towards $\pm\infty$ when approaching the singularity, and so will do the eigenvalues. Numerically, as proposed by Refs.\cite{McDonald13,Boccelli22}, one can prevent the closing flux of Eq.~\eqref{eq:sME}, becoming singular as $\beta\rightarrow0$. These references propose to limit the value of $\beta$ around the singularity. In our simulations, we fixed it to $\beta_\text{min} = 10^{-4}$, similar to what was done in Refs.\cite{McDonald13, Boccelli22}. The value of the cutoff $\beta_\text{min}$ did not seem to have a significant impact on the profiles, if small enough, but the time step becomes extremely small for smaller values of $\beta_\text{min}$ if some points of the simulation are close to the Junk line. As will be seen in section~\ref{sec:results}, under low pressure conditions the present of the singular Junk line becomes a problem.

\subsubsection{Hyperbolic quadrature method of moments (HyQMOM)}

The quadrature method of moments (QMOM) considers a VDF that is a sum of Dirac delta distributions (that represent the quadrature points to approximate integrals over the distribution) \cite{McGraw97}, formally,
\begin{equation}
    f_\mathrm{QMOM}(c) = \sum_{k=0}^N w_k \delta(c - c_k) \, .
\end{equation}
where $w_k$ are the weights of each Dirac delta distribution and $c_k$ is the abcissa of the Dirac delta distributions. 

The Hyperbolic QMOM (HyQMOM) closure \cite{Fox18} is a variant of QMOM that adds a condition that ensures the hyperbolicity of the system of equations. 
This closure will therefore consider $2N-1$ moments for $N$ Diracs.

In the case of a 5-moments system, we impose the fifth order moment to respect the following relation,
\begin{equation}
    s_\star = 2 r_\star q_\star - q_\star^3 \nonumber. 
\end{equation}
The closing flux is presented in Fig.~\ref{fig:realizability_eachMOM} along with the other considered closures.

The VDF will therefore consist of two Dirac distributions with abcissae that depend on the moments, and one Dirac fixed at the center of the VDF, namely at the drift velocity (i.e., $c=0$). Therefore, in our case, we will consider a VDF of the form
\begin{equation}
    f_\mathrm{HyQMOM}^\text{5M}(\nC) = \frac{\rho}{m \change{\vth}}\left[w_0 \delta(\nC) + w_1 \delta(\nC - c_1) + w_2 \delta(\nC - c_2)\right] \, .
\end{equation}
where the weights and abscissa can be analytically computed from the moments\cite{Fox18}, as follows,
\begin{eqnarray} \label{eq: HyQMOM inv}
    w_0 &=& 1 - (w_1 + w_2) \,,\quad\quad w_1 = \alpha w_2 \,,\quad\quad w_2 = \frac{\alpha}{(1+\alpha) c_1^2} \nonumber \\
    c_1 &=& \sqrt{\frac{r_\star}{1 - \alpha + \alpha^2}} \,,\quad\quad c_2 = - \alpha c_1 \nonumber \\
    \alpha &=& \frac{2r_\star - q_\star^2 - \lvert q_\star \rvert \sqrt{4r_\star - 3q_\star^2}}{2 (r_\star - q_\star^2)}.
\end{eqnarray}
Note that, in general, the QMOM closure for more than $5M$ does not allow for analytical expressions of the closing flux nor the VDF and they need to be computed through an eigenvalue problem with the Chebychev algorithm\cite{Marchisio13}.

\subsubsection{Extended QMOM (EQMOM)}

The Extended QMOM (EQMOM) closure \cite{Chalons10} is similar to the QMOM closure where the Dirac distributions are replaced by continuous functions, such as Gaussians with a single temperature. 

In the 5M case, the EQMOM VDF will therefore have the form,
\begin{equation}
     f_\mathrm{EQMOM}^\text{5M}(\nC) = \frac{n}{\change{\vth}}\left[\rho_1^\star\,\mathcal{M}_\star\left(\frac{\nC - c_1^\star}{a_\star}\right) + \rho_2^\star\,\mathcal{M}_\star\left(\frac{\nC - c_2^\star}{a_\star}\right)\right],
\end{equation}
where the normalized Maxwellians are defined in Eq.~\eqref{eq:NormalizedMaxwellian} and the normalized parameters are computed from the standarized moments as follows\change{:} \\
\change{Denoting $b_\star = 1 - a_\star^2$, the parameters are defined as
}\begin{eqnarray}
    && \change{c_1^\star = -\sqrt{\frac{\rho_2^\star}{\rho_1^\star} b_\star} \,,\quad\quad c_2^\star = \sqrt{\frac{\rho_1^\star}{\rho_2^\star} b_\star},} \nonumber \\
    && \change{\rho_1^\star = \frac{1}{2} + \frac{q_\star}{2 \sqrt{q_\star^2 + 4 b_\star^3}}} \,,\quad\quad \rho_2^\star = 1 - \rho_1^\star \,.
\end{eqnarray}
and \change{$b_\star$} is a solution of the equation
\begin{equation}
    \change{b_\star^3 + \frac{r_\star - 3}{2} b_\star - \frac{(q_\star)^2}{2} = 0,}
\end{equation}
that gives, using the third order polynomial formula
\begin{eqnarray} \label{eq: EQMOM b}
    && \change{b_\star = C_3 - \frac{C_2}{C_3}} \\
    && \text{with} \quad C_3 = \left(C_1 + \left(C_1^2 + C_2^3\right)^{1/2}\right)^{1/3} , \quad 
    C_1 = \frac{q_\star^2}{4} \quad \text{and} \quad C_2 = \frac{r_\star - 3}{6},
\end{eqnarray}
where the roots must be treated as complex principal root. 

\begin{figure}
    \centering
    \includegraphics[width=\linewidth]{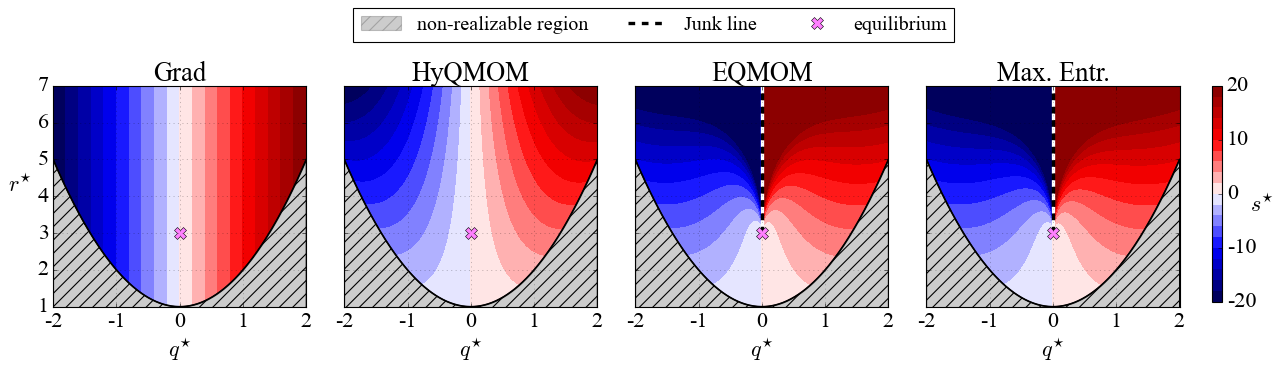}
    \caption{Variation of the closing flux $s_\star$ as the heat flux $q_\star$ and the kurtosis $r_\star$. The hatched zoned represent the non-realizable region where no positive function can have such heat-flux and kurtosis, and the dashed line represents the Junk line and the purple cross the thermodynamic equilibrium.}
    \label{fig:realizability_eachMOM}
\end{figure}

Finally, this closure allows for an analytical closing flux, written as,
\begin{eqnarray}
    s_\star &=& \frac{q_\star^3}{b_\star^2} + (10 - 8 b_\star) q_\star \, .
\end{eqnarray}
Like the maximum entropy closure, EQMOM has the same singular subspace, as shown in Fig.~\ref{fig:realizability_eachMOM}. 
Both closures are in fact very similar. 
The advantage of EQMOM is its mathematical simplicity compared to maximum entropy that requires the inversion of an integral system. 
However, as will be seen in  Sec.~\ref{sec:results}, the latter is better at capturing a wide variety of VDF forms. 
The singular subspace problem is addressed using the same approach as in maximum entropy.

In Fig.~\ref{fig:realizability_eachMOM}, we compare the standarized \change{hyperskewness} for the different models in the realizability space of the $5$M closures, where we represented the singular subspace for EQMOM and maximum entropy (Max. Entr.). We can clearly see that these two closures are indeed almost identical. Alternatively, Grad closure shows a great simplicity as the hyper skewness only depends on the heat flux. Finally, the HyQMOM closure has no singularities and a behavior that resembles Grad around the equilibrium, but that also depends on the kurtosis far from equilibrium.

\section{Numerical methods \label{sec:num}}

The presented moment closures are non-linear system of equations that can be written in the form:
\begin{equation}\label{eq:generalSystem}
    \frac{\partial \mathbf{M}}{\partial t} + \frac{\partial \mathbf{F}(\mathbf{M})}{\partial x} + \mathbf{B}(\mathbf{M})\frac{\partial\mathbf{M}}{\partial x} = \mathbf{S}(\mathbf{M}),
\end{equation}
where $\mathbf{M}$ are the velocity-moment variables, $\mathbf{F}(\mathbf{M})$ are the fluxes, $\mathbf{B}(\mathbf{M})$ is the non-conservative product matrix (which is $\mathbf{B}(\mathbf{M}) = 0$ for all the closures except for regularized Grad), and $\mathbf{S}(\mathbf{M})$ is the source term that contains the electric field and the collisional terms, the notations for the numerical scheme are given in Appendix \ref{sec:Notation}. The numerical resolution of Eq.~\eqref{eq:generalSystem} presents two main difficulties: (i) in high-order moment closures it is often difficult to have analytical approximations for the numerical fluxes, as required for computationally-efficient schemes, and (ii) the numerical treatment of this non-conservative product is technically complex and requires specific numerical schemes. In the following, we present a second-order numerical scheme that aims at providing efficient and stable solutions.

\subsection{Second-order finite volume scheme for Maximum Entropy, HyQMOM, and EQMOM closures}

The system of equations of Eq.~\eqref{eq:generalSystem}, with $\mathbf{B}(\mathbf{M}) = 0$, can be discretized with classical finite volume schemes \cite{LeVeque02,Toro09}. In the finite volume discretization, the domain is divided into $N_C$ cells with size $\Delta x_i$ for $i\in\{1,\,2,\cdots,\,N_C\}$, were the time evolution of the cell-averaged conservative variables in the cell i $\mathbf{M}_i$ follows the equation,
\begin{equation}\label{eq:spaceSemiDiscretization}
    \frac{d \mathbf{M}_i}{d t} + \frac{1}{\Delta x_i} \left(\mathbf{F}_{i+1/2}- \mathbf{F}_{i-1/2} \right) = \mathbf{S}_i,
\end{equation}
where the fluxes \change{$\mathbf{F}_{i\pm1/2}$} are evaluated in the cell interface $i\pm 1/2$. For the time discretization, in this work we use a simple forward Euler scheme, as we are interested in steady-state solutions and the convergence times are relatively fast. As a result, the discretized equation reads
\begin{equation}\label{eq:spaceTimeDiscretization}
    \mathbf{M}^{n+1}_i = \mathbf{M}^{n}_i - \frac{\Delta t}{\Delta x_i} \left(\mathbf{F}^n_{i+1/2}- \mathbf{F}^{n}_{i-1/2} \right) + \Delta t\mathbf{S}_i^n.
\end{equation}  
Here, the cell-averaged source term is approximated with the cell averages variables, i.e., $\mathbf{S}_i^n = \mathbf{S}(\mathbf{M}^n_i)$. Alternatively, in order to compute the fluxes at the cell interface, the considered closures have the advantage that the homogeneous system of equation (i.e., with $\mathbf{S}(\mathbf{M}) = 0$), is a hyperbolic non-linear system of equations (i.e., the Jacobian matrix $\partial\mathbf{F}/\partial \mathbf{M}$ is diagonizable with real eigenvalues). As a result, Godunov schemes and approximated Riemann solvers can be applied to the discretization of the fluxes. In the present work, we use a Rusanov scheme that reads as follows,
\begin{equation}\label{eq:Rusanov}
    \mathbf{F}^n_{i+1/2} = \tfrac{1}{2}\left[\mathbf{F}(\mathbf{M}^n_{L}) + \mathbf{F}(\mathbf{M}^n_{R})\right] - \tfrac{|\lambda^{max}_{i+1/2}|}{2}\left(\mathbf{M}^n_{R} - \mathbf{M}^n_{L}\right),
\end{equation} 
where the sub-index $R/L$ represents the reconstructed values on the right and left of the interface and $|\lambda^{max}_{i+1/2}|$ is the \change{local} maximum spectral radius of the Jacobian matrices with the left and right states. The reconstructed values of the conservative variables are computed with a Monotonic Upstream-centered Scheme for Conservation Laws (MUSCL) scheme \cite{vanLeer79} with a minmod limiter. We reconstruct the primitive variables $\mathbf{P}$ (defined in Appendix \ref{sec:Notation}) and compute the conservative variables \change{at} the interface from the reconstructed primitive variables. We stress that this choice in the reconstructed variables plays a very important role in the stability of the solution. We also tested the reconstruction of the conservative variables, which provides unstable solutions, in particular in the sheath region. 

The numerical scheme presented in Eq.~\eqref{eq:spaceTimeDiscretization} with \eqref{eq:Rusanov} has \change{the following stability condition }
\begin{equation}\label{eq:CFL}
    \text{CFL} = \min\left(\Delta t\, \underset{i\in N_c}{\max}\left(\frac{|\lambda^{max}_{i}|}{\Delta x_i}\right),\, \Delta t\nu_{ig}\right)<1.
\end{equation}

\subsection{Second-order finite volume scheme for the regularized Grad closure}

The regularization of Grad's model is proposed by Cai et al.~\cite{Cai14} such that the matrix $\partial \mathbf{F}/\partial\mathbf{M}+\mathbf{B}$ is hyperbolic.  This method is an elegant solution to the loss of hyperbolicity of original Grad's moment method but it comes with the drawback of adding a non-conservative product to the equations. A consequence of the non-conservative character of the fluxes is that classical Godunov's schemes cannot be applied as there \change{are} no general Rankine-Hugoniot relations. A solution to this difficulty was proposed by Dal Maso et al.~\cite{DLM} with a generalization of the weak solution introducing the notion of a ``path'' in the phase space of the variables \cite{Chalons17}. This ``path'' adds a degree of freedom to the numerical method. In the present work, we adapt the work of Dumbster \& Balsara \cite{Dumbser16} to our sytem of equations. The discretization of Eq.~\eqref{eq:generalSystem} is done by writing the fluctuation form of the discretized system, as follows,
\begin{equation}\label{eq:spaceTimeDiscretizationRegGrad}
    \mathbf{M}^{n+1}_i = \mathbf{M}^{n}_i - \frac{\Delta t}{\Delta x_i} \left(\mathbf{D}^-_{i+1/2}+ \mathbf{D}^{+}_{i-1/2} \right) - \frac{\Delta t}{\Delta x_i}\left(\mathbf{F}^-_{i+1/2}- \mathbf{F}^{+}_{i-1/2} \right)-\frac{\Delta t}{\Delta x_i}\hat{\mathbf{B}}(\mathbf{P}^n_i)\Delta\mathbf{P}^n_i + \Delta t\mathbf{S}_i^n,
\end{equation}  
where the details of the discretization of the fluxes and the non-conservative matrix are given in Appendix \ref{sec:Notation}. The stability condition of this scheme is the same as in Eq.~\eqref{eq:CFL}.

\subsection{Analytical and approximated expressions of the eigenvalues of the Jacobian of the flux }
In order to estimate $|\lambda^{max}_{i+1/2}|$, needed for the discretization of the fluxes, we need to evaluate the eigenvalues of the Jacobian matrix of the flux. This is one of the major difficulties in high-order moment models as they, in general, do not have analytical expressions. The numerical computation of $|\lambda^{max}_{i+1/2}|$ may result in computationally expensive numerical calculation of the Jacobian matrix and its eigenvalues. In order to avoid this, in this work, we provide analytical expressions (when possible) and interpolative expressions in order to have computationally efficient and accurate algorithms.

\paragraph{HyQMOM:}
The eigenvalues of the HyQMOM Jacobian flux can be analytically computed and read as follows\cite{Fox18},
\begin{equation}
    \lambda^{HyQMOM}_{0, 1, 3, 4} = u + \sqrt{\tfrac{p}{\rho}}\left(\tfrac{q_\star}{2}\pm\sqrt{1 + \Upsilon+\tfrac{{q_\star}^2}{4}\pm\sqrt{\Upsilon(1+\Upsilon)}}\right),~~~\text{and}~~~\lambda^{HyQMOM}_2 = u
\end{equation}
where $\Upsilon = r_\star - q_\star^2 -1$ can be proven to be a non-negative quantity for any realizable set of moments and, hence, the eigenvalues are indeed always real. 

\paragraph{Maximum entropy:} The eigenvalues of the maximum entropy closure cannot be computed analytically. In this work, we use the approximation proposed by Baradaran \cite{Baradaran15} (c.f. Boccelli\cite{Boccelli24} for the 3D case), which we give for completeness, as follows,
\begin{equation}
    \lambda^{Max. Ent.}_m \approx u + \change{\vth}\lambda^\star_m \quad \text{with} \quad m\in\{0,1,2,3,4\}
\end{equation}
and 
\begin{align}
    &\lambda^\star_{0,4}=\tfrac{{q_\star}\pm\sqrt{q_\star^2 \mp \tfrac{4}{5}q_\star^2\beta C_\star+4\beta^2 Y_\star}}{2\beta}\pm \tfrac{8}{10}C_\star, \quad\quad \lambda^\star_{1,3}=\tfrac{{q_\star}\pm\sqrt{q_\star^2 \mp \tfrac{6}{5}q_\star^2\beta C_\star+4\beta^2 X_\star}}{2\beta}\mp \tfrac{3}{10}C_\star,\nonumber \\
    &\lambda^\star_2=\tfrac{2q_\star(q_\star^2\sqrt{\beta}+2\beta^3)}{\sqrt{\beta^3}(q_\star^2+2\beta^2)}-\lambda^\star_0-\lambda^\star_1-\lambda^\star_3-\lambda^\star_4,
\end{align}
where 
\begin{align}
    & X_\star=A_\star+\left(\tfrac{3}{10}\right)^2C_\star^2+\tfrac{3}{5}C_\star\sqrt{A_\star}, \quad\quad  Y_\star=B_\star+\left(\tfrac{8}{10}\right)^2C_\star^2-\tfrac{8}{5} C_\star \sqrt{B_\star}, \quad\quad C_\star = \sqrt{3-3\beta}, \nonumber \\
    & A_\star=5-4\sqrt{\beta}-\sqrt{10-16\sqrt{\beta} + 6\beta}, \quad\quad B_\star=5+4\sqrt{\beta}-\sqrt{10-16\sqrt{\beta} + 6\beta}.
\end{align}

\paragraph{EQMOM:}
The eigenvalues of EQMOM cannot be computed analytically. In the present work, we \change{provide} an interpolative approximation of the spectral radius, similar to the method of Baradaran\cite{Baradaran15} for maximum entropy shown above. 
The asymptotes of the highest eigenvalues at positive and negative high heat-flux are respectively $q_\star/b_\star + \sqrt{3(1-b_\star)}$ and $\sqrt{3(1-b_\star)}$. 
The asymptotes of the lowest eigenvalues at positive and negative high heat-flux are respectively $-\sqrt{3(1-b_\star)}$ and $q_\star/b_\star - \sqrt{3(1-b_\star)}$. As a result, the spectral radius can be estimated with an interpolation between these two asymptotes, as follows,
\begin{equation}
    |\lambda^{\star,HyQMOM}_{max}| \approx \max\left(\tfrac{1}{2} \left|\tfrac{q_\star}{b_\star} + \sqrt{4 + \left(\tfrac{q_\star}{b_\star}\right)^2} + \sqrt{3(1 - b_\star)}\right|,\,
    \tfrac{1}{2} \left| \tfrac{q_\star}{b_\star} - \sqrt{4 + \left(\tfrac{q_\star}{b_\star}\right)^2} - \sqrt{3(1 - b_\star)}\right|\right).
\end{equation}
This fit is able to approximate the spectral radius $\lambda^{HyQMOM}_{max} \approx u + \change{\vth}|\lambda^{\star,HyQMOM}_{max}|$ within a 10\% accuracy in the worst case, which is shown to be valid for the studied cases. 

\paragraph{Globally hyperbolic regularized Grad}

The advantage of the regularization proposed by Cai et al.~\cite{Cai14} is that the system has an analytical solution for the eigenvalues at arbitrary moment order. In our case, they read as follows,
\begin{equation}
    \lambda^{Reg.~Grad}_{0, 1, 3, 4} = u\pm \change{\vth}\sqrt{5\pm\sqrt{10}}\quad\text{and}\quad\lambda^{Reg.~Grad}_2 = u.
\end{equation}

\subsection{Ionization and electric field}

In this work, we focus on the  the influence of the different moment models on the ion dynamics at different pressure regimes. Following the methodology proposed by Bocelli\cite{Boccelli20,Boccelli20-HET}, we will impose the quantities that depend \change{on} the electron dynamics. These are the ionization rate $S_{\text{iz}}$ and the electric field $E$, that will be taken from the PIC results and the values will be interpolated to the mesh used for the moment models. In this way, the accuracy of the different ion moment models can be assessed, independently of the model used for the electrons. A self-consistent simulation where the two charged species are coupled is also possible, as previously done in Refs.~\cite{Alvarez20, Gangemi25}, and is left for future work comparing electron moment closures.

\subsection{Numerical grid, boundary and initial conditions}

The simulations are run on a non-uniform grid with a cell size of $0.5$~mm in the bulk that decrease to $10$~$\mu$m in the sheath. The simulations are initialized with a constant background of density $n = 10^{15}$ m$^3$ with a Maxwellian distribution at rest with the gas temperature $T = 0.025$ eV. The boundary conditions are imposed with a ghost cell strategy with outflow conditions, where we assume that the ion flow is supersonic at the boundary.
The CFL was set to $0.1$ at the beginning due to the large variations during the transient and can then be increased to nearly $1$.

\section{Results \label{sec:results}}

We present here the results of the high-order moment simulations representing the ions in the numerical setup of an argon plasma between two floating walls (described in Section \ref{sec:NumericalSetUp}) and compare them with the PIC-MCC simulations with the conditions summarized in Table \ref{tab:NumParams}. 
As discussed earlier, the simulations study different pressure regimes. We vary the pressure of the background gas from $p_g = 10$~Pa, that corresponds to an ion-neutral mean free path much smaller than the domain, i.e., $\lambda_{mfp}/L = 0.01$, to $p_g = 0.01$~Pa, that corresponds to an ion-neutral mean free path larger than the domain, i.e., $\lambda_{mfp}/L = 10$. We will therefore refer to these two pressures as respectively high and low pressure in the rest of the paper, and the other two pressures as intermediate pressures. 

In order to decrease the noise of the PIC data, the results have been averaged over $3.6\cdot10^{6}$ time steps, after convergence which takes of the order of $10^7$ time steps. The high-order moment equations are solved in time until convergence, which requires of the order of $10^5$ to $10^6$ time steps (depending on the pressure regime and the closure). Furthermore, moment results do not require to be time averaged as they are free of statistical noise. 
The high-order moments models will also be compared to several classical fluid models. One based on the isothermal fluid equations (i.e. 2M) and two non-isothermal fluid models (with 3M) based on different closures: the isotropic Maxwellian and an anisotropic Maxwellian (i.e. with two temperatures, one in the direction $x$ and another in the perpendicular direction considered to be equal to the \change{neutral} gas temperature).

\subsection{Moments profiles}

\begin{figure}
    \centering
    \includegraphics[width=\linewidth]{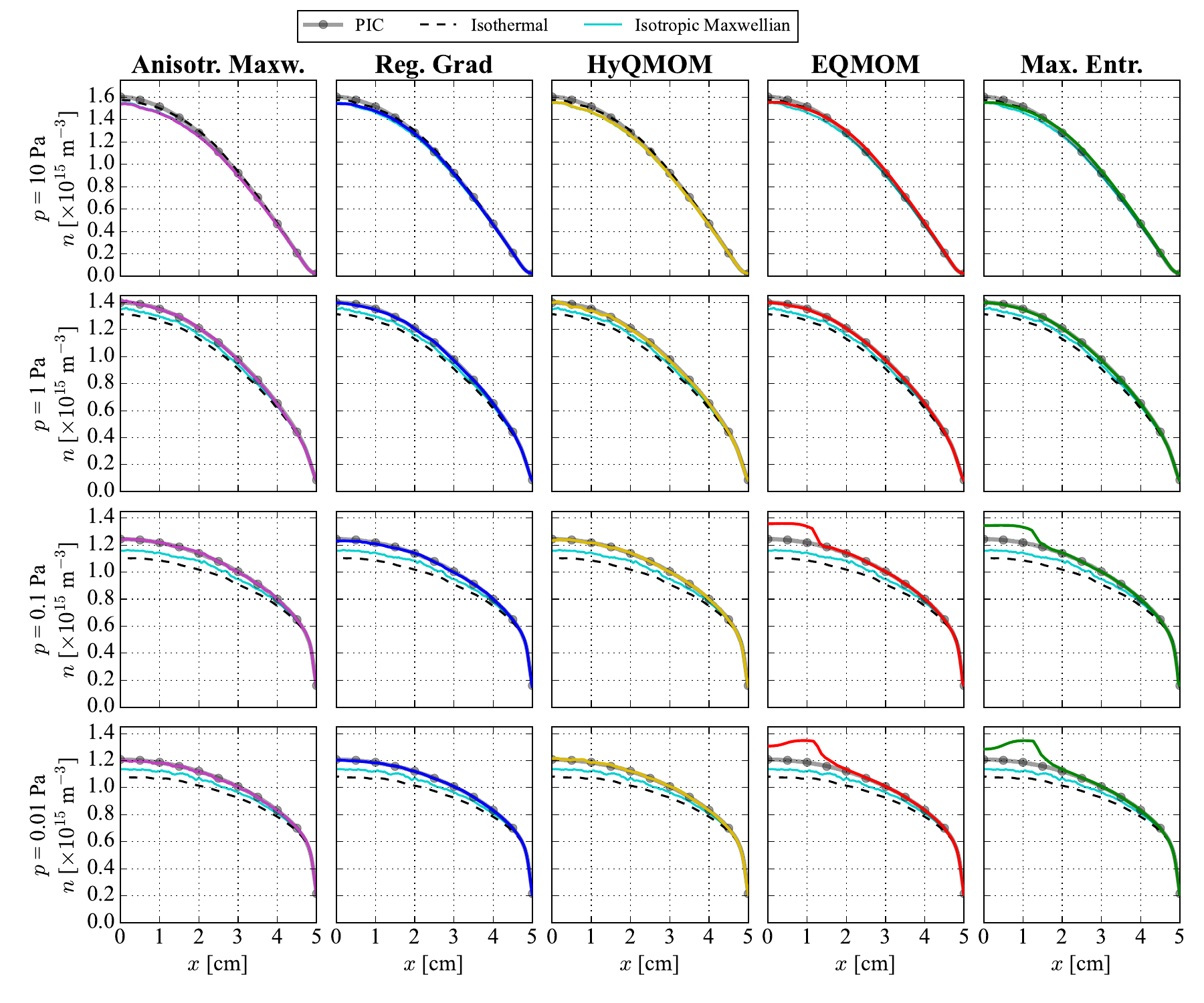}
    \caption{Density profiles of all the simulations for four pressures.}
    \label{fig:n_profiles}
\end{figure}

\begin{figure}
    \centering
    \includegraphics[width=\linewidth]{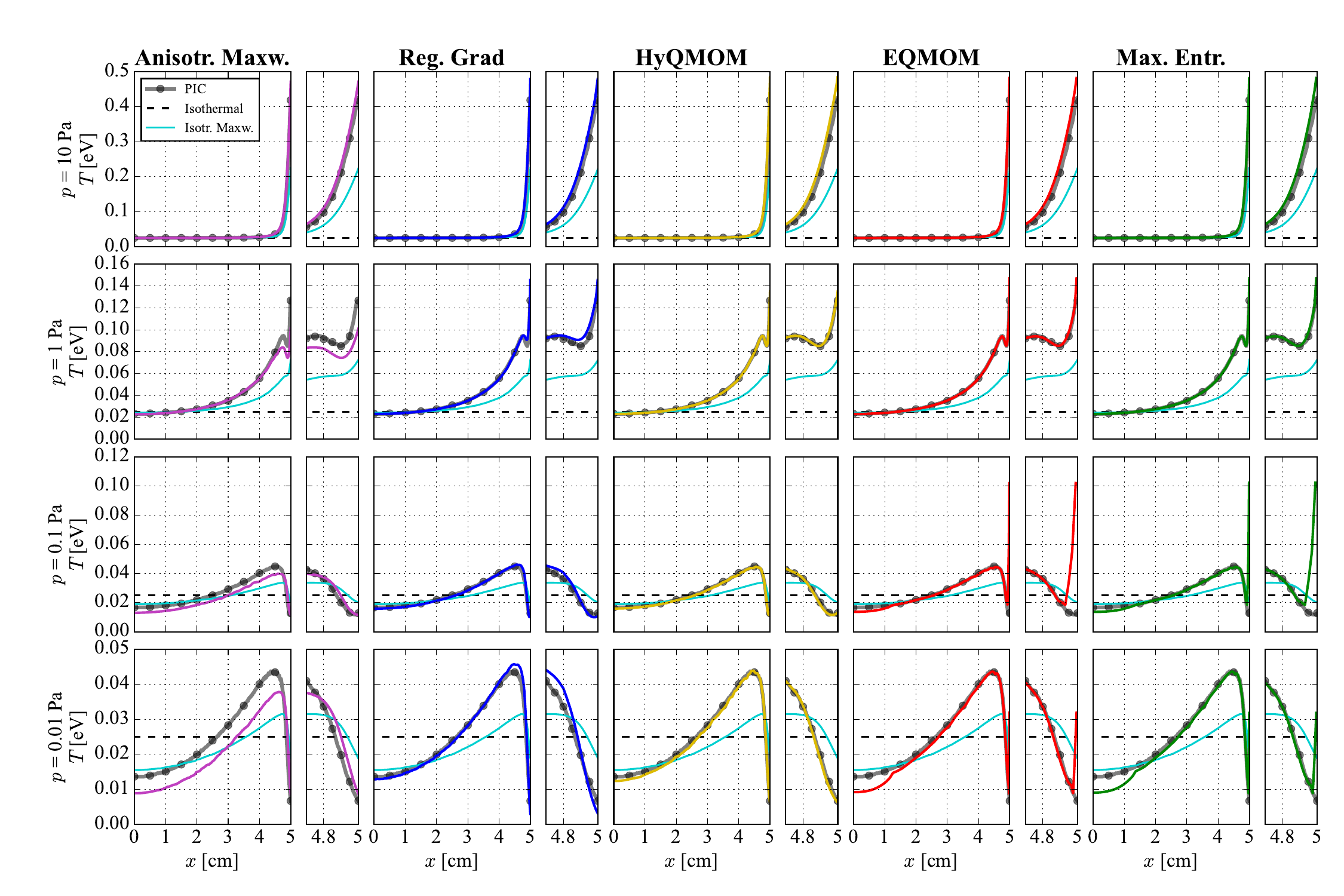}
    \caption{Temperature profiles of all the simulations for four pressures. For each plot we show a zoom of the sheath next to it.}
    \label{fig:T_profiles}
\end{figure}

We first present the moment profiles for every model considered at the studied gas pressures. 
Due to the symmetry of the system, we show only half of the domain for each profile. 
In Fig.~\ref{fig:n_profiles}, we present the density profiles for the 5M models as well as the 3M model with anisotropic Maxwellian compared to the PIC simulations as well as with the isothermal and 3M isotropic Maxwellian models. As it is shown in the figure, the density profile varies from a dome-like profile (Shottky collisionally dominated regime) at high-pressure to a flat profile with a sharp drop in the sheath (Tonks-Langmuir nearly collisionless regime), at low-pressure. 
In the density profiles, we can see that, in general, regularized Grad (Reg.~Grad) and HyQMOM are able to capture almost perfectly the density profile at all pressures. 
However, both EQMOM and maximum entropy (Max. Entr.) present a spurious density gradient in the center of the domain at low pressure. 
This feature is associated with the singularity of the closing flux (Junk line) when the kurtosis is positive and the heat-flux is null. 
As compared to the isothermal 2M and isotropic Maxwellian 3M moments, the moment models improve the representation of the density profile in the low and intermediate pressures, whereas they compare similarly at high pressures. 
Alternatively, the anisotropic Maxwellian 3M model is able to largely improve the density profile, although, as it will be shown later, is not able to correctly predict the higher-order moments, the heat-flux at the wall nor the VDF.

The temperature (the $xx$ component of the temperature tensor) profiles are shown in Fig.~\ref{fig:T_profiles}. Note that a zoom in the sheath is presented on the right of each profile. At high pressure, the temperature is constant at $T_\text{g}$ except in the sheath region, where it increases because of the effect of the strong electric field and the collisions with the gas. 
Alternatively, the temperature at low and intermediate pressures is lower than the gas temperature in the center (because of the strong advection) and larger in the sheath (because of the electric field). 
As in the density profiles, the high-order moment models, in particular HyQMOM, are able to capture with very high fidelity the temperature profiles both in the bulk and in the sheath at all pressures. 
We also see the effect of the Junk line at low pressure for EQMOM and maximum entropy, where the temperature decreases slightly more than expected in the center of the domain. 
Note that the large increase at the wall at low pressures for these closure is a consequence of lack of resolution. This is due to the lack of stability and robustness of these models (because of the Junk line) that prevents to run these simulations with better resolution (because the time step tends to zero due to the presence of the Junk line). 
Finally, the 3M and 2M classical fluid models are less accurate in most of the domain at all pressures. The 3M anisotropic Maxwellian model is as good as the moment models at high pressure, but loses its accuracy as the pressure decreases. This is because of the lack of heat-flux, which becomes important at low pressures, as it will be shown below. The isotropic 3M model has the right qualitative behavior but lacks of precision at all pressures. 

\begin{figure}
    \centering
    \includegraphics[width=\linewidth]{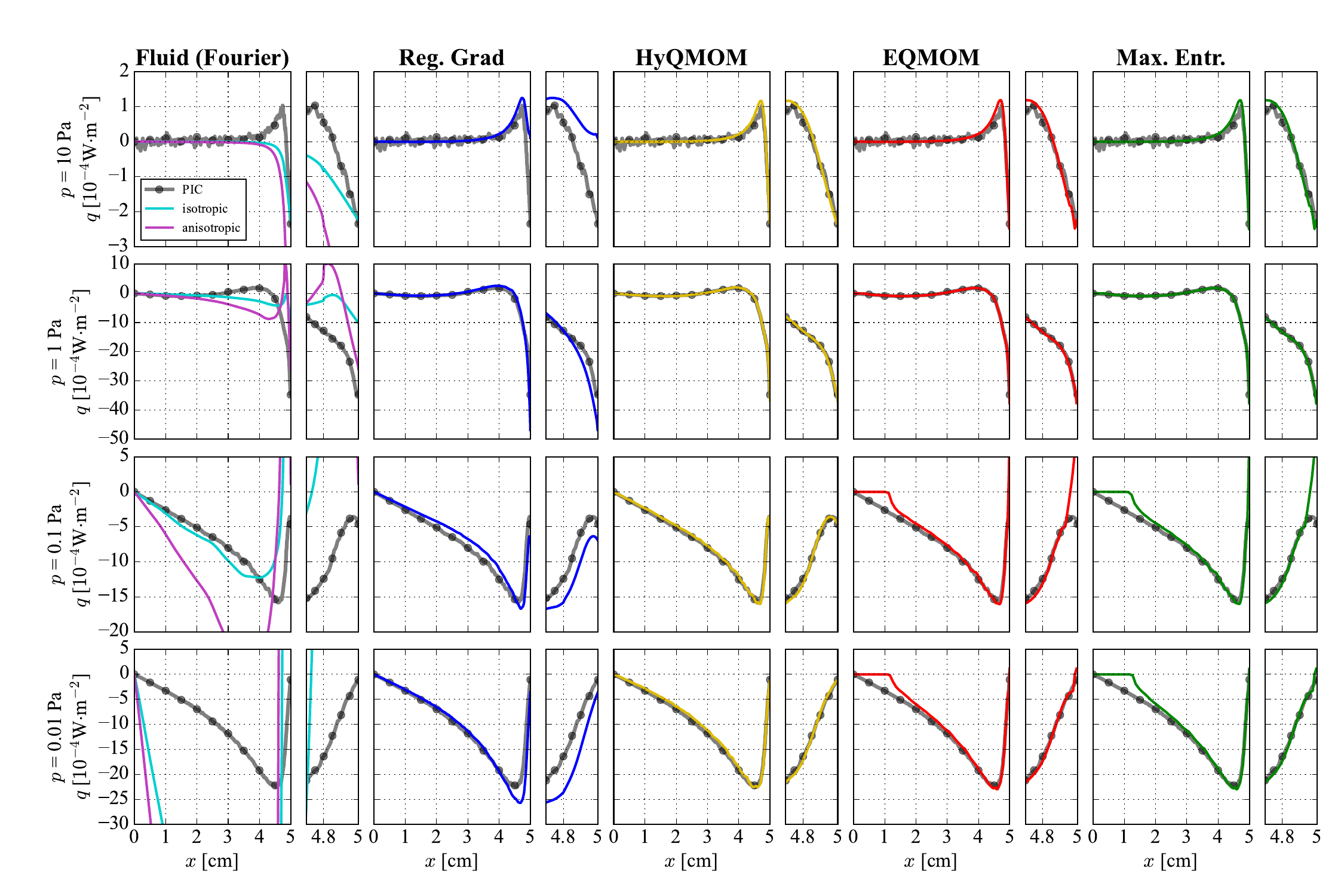}
    \caption{Heat flux profiles of all the simulations for four pressures. We represent the Fourier heat flux of the non-isothermal Maxwellian cases in dashed lines. For each plot we show a zoom of the sheath next to it.}
    \label{fig:q_profiles}
\end{figure}

The real advantage of the high-order moment models is their ability to capture higher-order moments without statistical noise that is characteristic of particle-based kinetic simulations. We show the heat flux profile in Fig.~\ref{fig:q_profiles}. At high pressure, the heat-flux is almost zero in the bulk and it presents a sharp drop in the sheath, whereas it is large in the bulk for low pressures. 
The four moment models overall capture the kinetic behavior with a high level of accuracy. However, we still see the problem of the Junk line in the center of the domain for EQMOM and maximum entropy. 
We also see small discrepancies in the sheath for \change{Reg.}~Grad, which can be explained due to the lack of precision of the moment of order $4$. 
For comparison, we also present the heat-flux obtained with the Fourier approximation\cite{Kremer10}
\begin{equation}
    Q_{xxx} = \tfrac{6}{5} q_x = - 3 \frac{p k_\text{B}}{m \nu} \partial_x T \,,
\end{equation}
applied to 3M fluid models with the isotropic Maxwellian and the anisotropic Maxwellian. 
We clearly see that the Fourier approximation is not well-suited here and that it almost always overestimates (in norm) the heat-flux, especially at low pressure. This justifies a high-order moment approach in order to capture the heat-flux to the wall in low-temperature plasmas: Fourier law is not able to represent the heat-flux inside the sheath due to the large deviation from thermodynamic equilibrium.

\begin{figure}
    \centering
    \includegraphics[width=\linewidth]{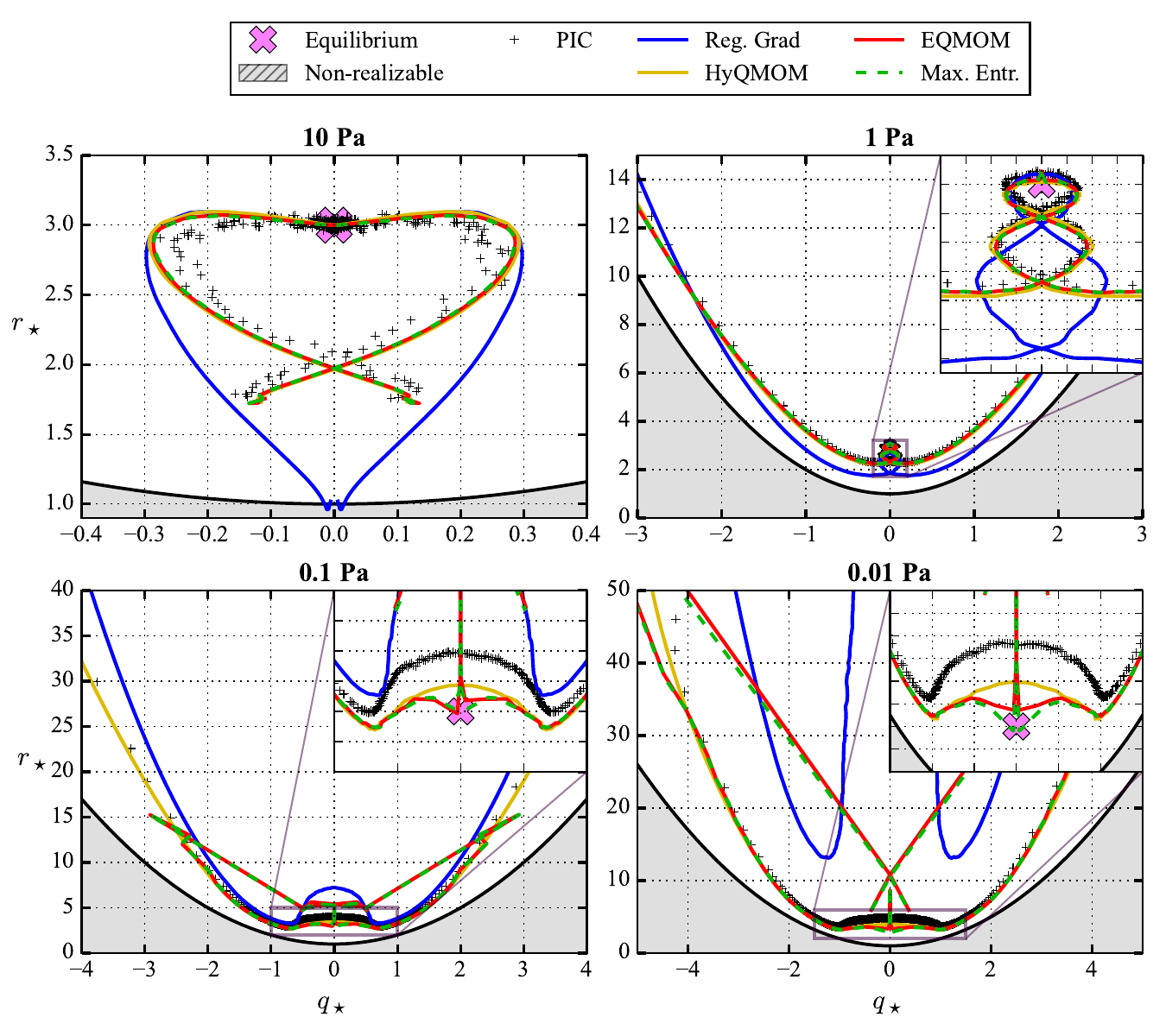}
    \caption{High-order moments profiles plotted in the realizability space (heat-flux - kurtosis space) for four pressures.}
    \label{fig:realizability}
\end{figure}

In figure \ref{fig:realizability} we show the profiles of the standardize heat-flux and kurtosis represented in the phase-space (the so-called realizability space). 
At high pressures (upper plots), the VDF is closer to a Maxwellian in the bulk, so the heat flux and kurtosis stay mostly around the equilibrium point ($q_\star = 0$ and $r_\star = 3$, represented with a pink cross). 
At these high pressures, all the moment models capture the kinetic behavior with great accuracy in the bulk. 
In the sheath, while the moments deviate greatly from equilibrium, all except \change{Reg.}~Grad still match the kinetic behavior with a very high level of accuracy. 
In \change{Reg.~Grad's simulations}, the moments even leave the realizability region at the boundaries, in the upper left plot, as a consequence of a VDF that is not ensured to be positive by this method. 
On the other hand, at lower pressures (lower plots), HyQMOM is the most accurate overall, despite an offset of kurtosis in the bulk. 
For EQMOM and Maximum Entropy, we see clearly the problem of the crossing of Junk's line in the center of the domain. 
These two models exhibit an interesting behavior, where the moments peak in the Junk line after being deviated towards the equilibrium. 
A similar behavior can be seen in the results presented by Boccelli\cite{Boccelli22}. 
This nonphysical behavior is due to the artificial cutoff of the closing moment around Junk's line to avoid its divergence. 
Despite this problem in the center, EQMOM and maximum entropy are still highly accurate far from the center of the domain except in the last mesh points of the domain close to the wall, due to the lack of resolution of these two simulations at low pressures as mentioned in the previous paragraph.
Finally, \change{Reg.}~Grad highly overestimates the kurtosis everywhere. This overestimation is due to the regularization terms in the kurtosis equation, as mentioned in the previous paragraph. 
An interesting point to stress on is that despite the very large discrepancy of \change{Reg.}~Grad in the kurtosis, this has a minor impact in the previous moments (only small inaccuracies in heat-flux, as discussed above).

\subsection{VDF reconstruction}

We will now assess the accuracy of the different closures to reconstruct the VDF from the simulated moments. 
We stress the fact that, in order to obtain low-noise VDF from PIC simulations, the data were averaged over $3.6\cdot10^{6}$ time steps. Alternatively, the VDFs of the high-order moment models are instantaneous values, obtained with a method that is computationally almost as efficient as classical fluid models.

First, since the HyQMOM VDF consists of a sum of Dirac delta distributions, this model will not recover a continuous VDF. 
This is a major drawback if one is interested in the VDF shape, in particular for an accurate computation of the collisions or wall processes (e.g., sputtering). 
However, the three Diracs represent in fact quadrature points used to calculate integrals of a continuous distribution. 
In that sense, Fox et al.\cite{Fox23} developed a method, called generalized QMOM (GQMOM), to dramatically increase the number of Diracs while keeping the same first moments. The VDF can therefore be refined as much as desired, making it quasi-continuous. This method will be used in order to represent the reconstructed VDFs of the HyMOM results. 

\begin{figure}
    \centering
    \includegraphics[width=0.95\linewidth]{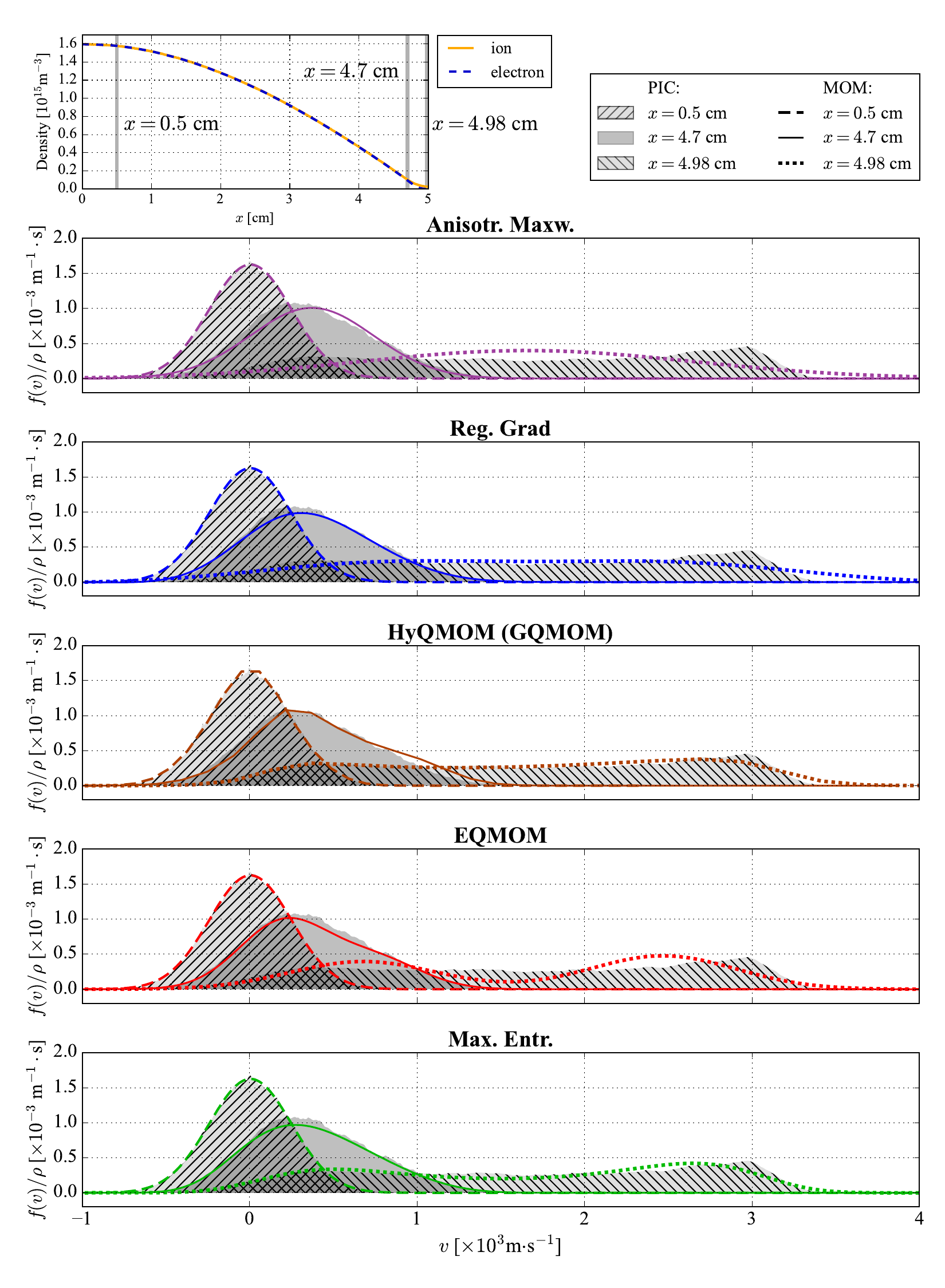}
    \caption{VDF from the kinetic simulations and reconstructed from the 5-moment simulation results for a pressure of 10~Pa. The upper plot shows the the positions at which the VDF are taken.}
    \label{fig:VDF_PIC&5M_HP}
\end{figure}

The results of the VDF reconstruction at high pressure ($p_g = 10$ Pa) are shown in Fig.~\ref{fig:VDF_PIC&5M_HP}, in three different points of the domain, including the bulk, the entrance of the sheath and close to the wall. 
The VDF is very close to Maxwellian in most of the bulk and can therefore be reproduced with great accuracy by the four 5M models in the bulk as well as the 3M model with anisotropic Maxwellian model. In the entrance of the sheath, the distribution has a positive skewness, which is captured by all the 5M models, and it is beyond the 3M representation. Close to the wall, the VDF is strongly non-Maxwellian, and the global shape of the VDF is recovered by all the models with remarkable accuracy, despite some small differences for EQMOM. 
We note small differences with the kinetic simulation in order to capture the fast decreasing tail on the high-velocities side. 

\begin{figure}
    \centering
    \includegraphics[width=0.95\linewidth]{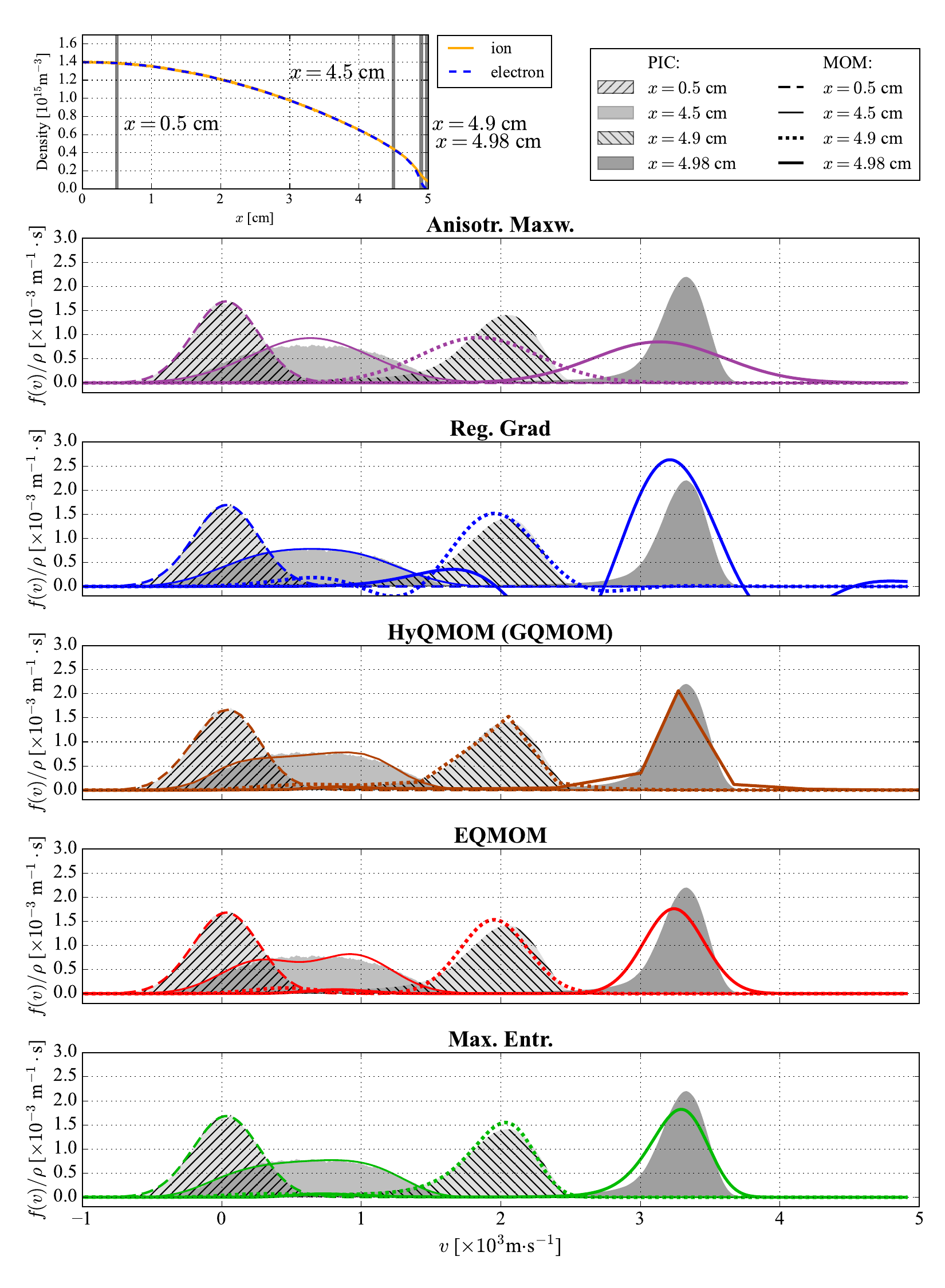}
    \caption{VDF from the kinetic simulations and reconstructed from the 5-moment simulation results for a pressure of 1~Pa. The upper plot shows the the positions at which the VDF are taken.}
    \label{fig:VDF_PIC&5M_IP}
\end{figure}

The intermediate pressure regime ($p_g = 1$ Pa) is presented in Fig.~\ref{fig:VDF_PIC&5M_IP}. We present the VDF in four points: in the center, in the presheath, in the entrance of the sheath, and close to the wall. The kinetic simulation shows that the VDF is nearly Maxwellian at the center, whereas it is very elongated in the presheath and it is a beam with a long tail in the sheath and close to the wall. As seen in the results, these VDFs are beyond the 3M description with an anisotropic Maxwellian, which provides a particular wrong representation in the sheath and close to the wall. Alternatively, all the 5M models largely improve the 3M results. We can see that the regularized Grad model has positivity problems (i.e., negative tails) in the sheath. Alternatively, maximum entropy provides an excellent representation at this pressure. EQMOM and GQMOM capture the right trend while presenting some differences, but still a remarkable accuracy.

\begin{figure}
    \centering
    \includegraphics[width=0.95\linewidth]{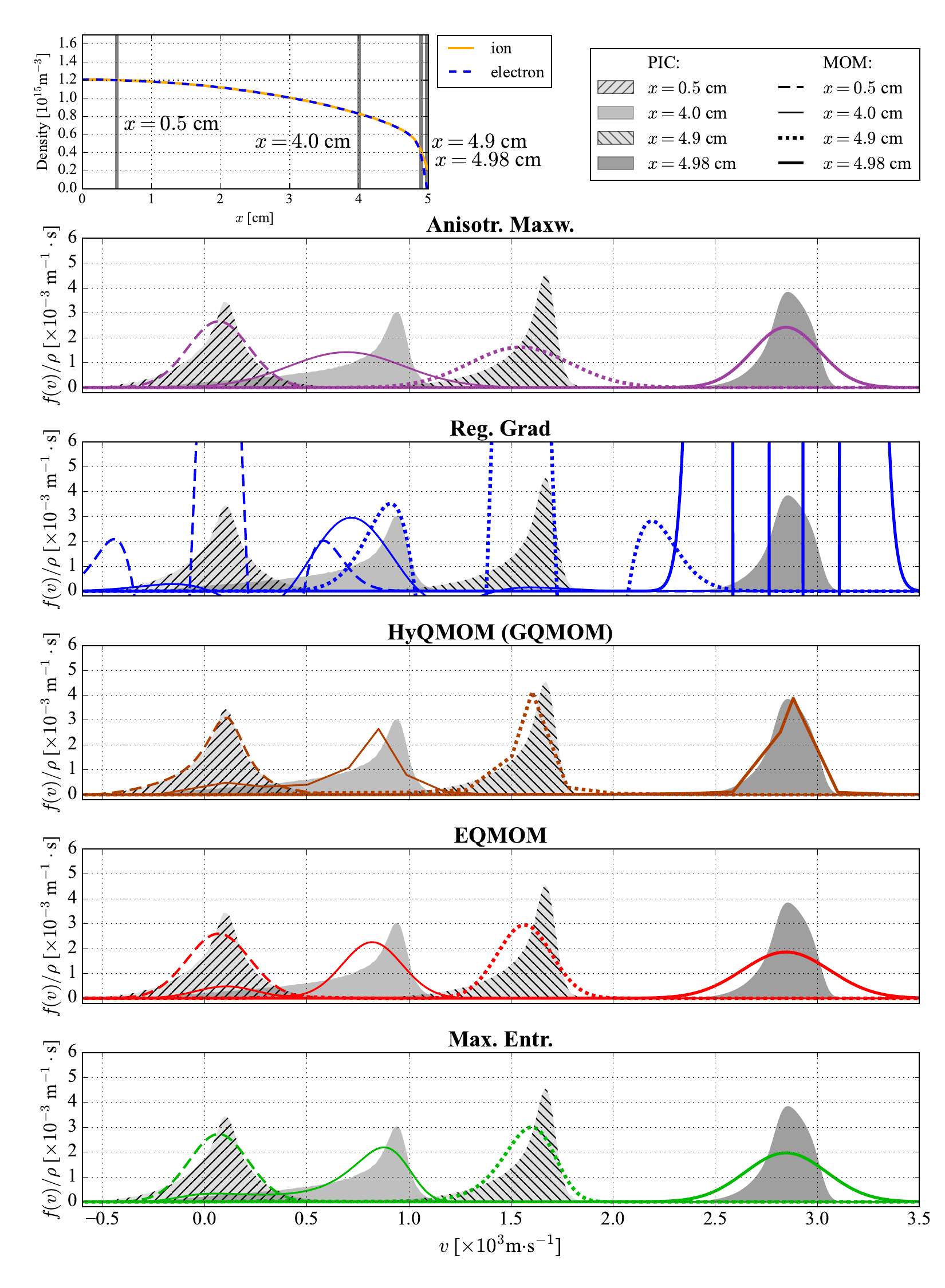}
    \caption{VDF from the kinetic simulations and reconstructed from the 5-moment simulation results for a pressure of 0.01~Pa. The upper plot shows the the positions at which the VDF are taken.}
    \label{fig:VDF_PIC&5M_LP}
\end{figure}

Finally, the low pressure ($p_g = 0.01$ Pa) is presented in Fig.~\ref{fig:VDF_PIC&5M_LP}. As in the previous case, we present the VDF in four points: in the center, in the presheath, in the entrance of the sheath, and close to the wall. In this case, the VDF in not Maxwellian in any of them. In the center of the domain, it presents heavy tails with a symmetric distribution, whereas in presheath, sheath and close to the wall, the distribution is very skewed, with a very long tail towards low energies. As a result, these very non-equilibrium distributions can be seen as a challenge for the different high-order moment models. As previously discussed, due to the presence of the Junk line, the moment inversion of maximum entropy and the EQMOM do not necessarily converge around the center of the domain and for visualization purposes are set to be Maxwellian. 
In the rest of the domain, these two closures still manage to capture the very long tail on the low-energy side of the VDF, but they lack precision in capturing the shape of the peak of the VDF. Nevertheless, they still give a better approximation than the anisotropic Maxwellian. 
\change{Reg.}~Grad is clearly far from its region of validity in the whole domain with VDFs that exhibit large negative parts. 
Alternatively, GQMOM is the only method that manages to capture a very accurate shape of the VDF in the center of the domain as well as a good approximation of the VDF in the presheath, sheath and close to the wall. In summary, the 5M HyQMOM (combined with GQMOM for the VDF reconstruction) shows a great ability to represent non-equilibrium distributions at all pressure both in the bulk and the sheath. 

\section{Conclusion and discussion \label{sec:conclu}}

We compared the numerical resolution of some of the most common high-order moment closures to kinetic simulations in a one-dimensional bounded low-temperature argon plasma. For each moment closure, we used expressions that allow for an efficient computation of the closing flux and a stable and accurate numerical scheme that is able to have a computational time that is comparable to classical 3M fluid models, but with a largely improved accuracy. The accuracy of each method to recover the profiles of fluid quantities (i.e., the density, the temperature) is similar for the four 5M models and always more accurate than classical 3M fluid models, except in the center of the domain at low pressure for EQMOM and maximum entropy, due to the singular behavior of these closures. In particular, the high-order moment closures are able to capture the ion temperature evolution in the sheath better than models assuming anisotropic Maxwellians, showing that the heat-flux in the sheath has an impact that cannot be modeled with simplified a Fourier-type heath-flux. Indeed, the high-order moment models are able to capture the heat-flux profile with high accuracy both in the bulk and the sheath, largely improving the classical Fourier approximation. We also compared the ability of each method to reconstruct the VDF from the first five moments, showing promising results.  At high pressure, all 5M closures recover the kinetic VDF with high fidelity. At low pressures, \change{Reg.}~Grad presents strong negative tails, EQMOM and maximum entropy have an improved fidelity as compared to a Maxwellian approximation and GQMOM presents a good accuracy both in the bulk and the sheath of the domain at all pressures.

We propose a summary of the advantages, drawbacks, and possibility of extension of each of the considered closures:
\begin{itemize}
    \item \change{Reg.}~Grad is the simplest method despite the difficulty associated to the numerical discretization, as shown in this work.  In our comparison to the kinetic results, \change{Reg.}~Grad's closure presents a good accuracy in the representation of moments up to the heat-flux in both the sheath and the bulk. However, it presents problems to capture the kurtosis, due to the error in the closing flux associated to the non-conservative terms of the regularization. As shown in the VDF reconstruction, at low pressures the tail of the VDF presents strong negative parts. In our work we present the 5M model, but extending \change{Reg.}~Grad to a larger number of moments is straightforward, with a moderate increase of the computational cost. In addition, Grad's closure allows for an analytical integration of the collisional terms for general collisional interactions.
    \item The maximum entropy closure shows some clear advantages in accuracy at high and intermediate pressures, capturing with great accuracy the transport and the kinetic VDF in all the domain. In the 5M case considered here, we used an interpolative method that shows to be computationally very efficient. However, the presence of the singular Junk line is a real drawback at low pressure, both for the accuracy and the stability of the method. The extension to higher-order moment closures is not straightforward and would require specific interpolative expressions. In addition, there will always be a singular subspace (due to the fact that the last coefficient of the maximum entropy distribution function must necessarily be negative), whose impact is difficult to foresee. Furthermore, another drawback is the high computational cost of the VDF reconstruction. As a result, the computation of the collisional terms for general interaction potentials adds a layer of complexity to this problem since there is no analytical simplification of the collision terms and a numerical integration would be required in a general case, loosing numerical efficiency. 
    \item The main advantage of EQMOM is that its closure is very similar to maximum entropy but with a much lower computational cost to reconstruct the VDF. In our results, we show that the reconstructed VDF is, however, slightly less accurate than the maximum entropy closure. Nevertheless, an increase in the number of moments considered is rather simple and one can use some efficient moment inversion algorithms to limit the computational cost. In addition, the computation of the collisional terms might be possible in an analytical manner as the model is based on Maxwellian distributions. 
    \item Finally, HyQMOM was found to be the most efficient, robust, accurate, and stable method that captures with very high fidelity all the simulated moments both in the sheath and the bulk at all the studied pressures. The VDF reconstruction is more challenging, but the GQMOM shows to be a great tool, allowing to capture VDFs with high fidelity to the PIC ones at all pressure regimes and both in the bulk and the sheath. In addition, as compared to PIC they do not have any statistical noise. Furthermore, as in the case of EQMOM, an increase in the number of moments is rather simple and there exists efficient algorithms, such as the Wheeler algorithm\cite{Wheeler74}, for the VDF reconstruction. In addition, the computation of the collisional terms for general collisional cross sections has analytical expressions.
\end{itemize}

These results prove that the use of high-order moment hierarchies to simulate ions in a low-temperature plasma is a sound approach and could allow an accurate modeling at a much lower cost than kinetic simulations. In the 5M moment cases shown here, the computational cost is comparable to this of classical 3M fluid models. The collision modeling with general collision cross sections is still a challenge in high-order moment closures. Some of the moment methods studied in this work (e.g., Grad, EQMOM, HyQMOM) allow for affordable collision computations that will be studied in a future work.

\begin{acknowledgments}
This work is part of a PhD funded by the Institut Polytechnique de Paris (IPP) and EUR PlasmaScience.
\end{acknowledgments}



\appendix

\section{Moment inversion in maximum entropy closures}\label{sec:MEInversion}
We solve the moment inversion system by performing the Newton-Raphson algorithm: writing the system as
\begin{eqnarray}\label{eq_Newton}
    F(U) = M \, ,
\end{eqnarray}
where $U = (k_0,..., k_4)$ is the vector of parameters of the VDF and $M = (1, 0, 1, q_\star, r_\star)$ the vector of normalized moments, 
one performs iteratively until convergence the following operation:
\begin{eqnarray}
    U_{n+1} &=& U_n - J^{-1} (F(U_n) - M) \,,
\end{eqnarray}
where $J = \partial F / \partial U$ is the Jacobian of the function $F$. \\
However, the convergence of this method is not ensured and requires an initial guess very close to the solution for it to convergence in a reasonable number of iterations. 
For this reason, instead of solving directly equation \ref{eq_Newton}, we solve iteratively the equation
\begin{eqnarray}
    F(U_k) &=& M_0 + (M - M_0)\frac{k}{N} \,,
\end{eqnarray}
where $M_0 = (1,0,1,0,3)$ is the equilibrium moment, and $N$ is chosen so that the convergence is fast enough. 
We found $N = \lfloor10 \,\text{max}(q_\star, r_\star - 3)\rfloor + 1$ to work most of the time.

\section{Notation for the numerical schemes}\label{sec:Notation}
\subsection{Definition of vectors and matrices}\label{sec:Notation1}
The velocity moments are
\begin{equation}
    \mathbf{M} = \left(\begin{array}{c}
         M^{(0)}  \\
         M^{(1)}  \\
         M^{(2)}  \\
         M^{(3)}  \\
         M^{(4)}  \end{array}\right) = 
         \left(\begin{array}{c}
         \rho  \\
         \rho u \\
         \rho u^2 + p  \\
         \rho u^3 + 3 u p + q  \\
         \rho u^4 + 6 u^2 p + 4 u q + r  \end{array}\right).
\end{equation}
We define the array of primitive variables, as follows,
\begin{equation}
    \mathbf{P} = \left(\begin{array}{c}
         \rho \\
         u \\
         p  \\
         q  \\
         \mathcal{K}  \end{array}\right) \,,
\end{equation}
where the kurtosis is defined as $\mathcal{K} = r - 3p^2/\rho$. 

The conservative flux reads
\begin{equation}
    \mathbf{F}(\mathbf{M}) = \left(\begin{array}{c}
         M^{(1)}  \\
         M^{(2)}  \\
         M^{(3)}  \\
         M^{(4)}  \\
         \rho u^5 + 10 u^3 p + 10 u^2 q + 5ru + s
    \end{array}\right),
\end{equation}
where the closing flux $s$ depends on the particular model. 
The Jacobian matrix of the transformation from conservative to primitive variables reads
\begin{equation}
    \frac{\partial\mathbf{M}}{\partial\mathbf{P}}  = \left(\begin{array}{c c c c c}
          1 & 0 & 0 & 0 & 0  \\
          u & \rho & 0 & 0 & 0  \\
          u^2 & 2\rho u & 1 & 0 & 0  \\
          u^3 & 3\rho u^2 + 3p & 3u & 1 & 0  \\
          u^4-\tfrac{3p^2}{\rho^2}& 4\rho u^3 + 12pu + 4q & 6u^2+6\tfrac{p}{\rho} & 4u & 1  \\
    \end{array}\right).
\end{equation}
The non-linear source term contains the electric field and the collisional terms, as follows,
\begin{equation}
\mathbf{S}(\mathbf{M}) = \left(\begin{array}{c}
         S_\text{iz}  \\
         M^{(0)}\tfrac{e E}{m}+ \mathcal{C}^{(1)}  \\
         2 M^{(1)}\tfrac{e E}{m}+ \mathcal{C}^{(2)} + S_\text{iz} \frac{k_\text{B} T_\text{g}}{m}  \\
         3 M^{(2)}\tfrac{e E}{m}+ \mathcal{C}^{(3)} \\
         4 M^{(3)}\tfrac{e E}{m}+ \mathcal{C}^{(4)} + 3 S_\text{iz} \left(\frac{k_\text{B} T_\text{g}}{m}\right)^2
    \end{array}\right).
\end{equation}
Finally, the non-conservative product matrix $\mathbf{B}$ is zero for all the models except for the regularized Grad. In that case, the matrix in conservative variables reads, 
\begin{equation}
    \mathbf{B}(\mathbf{M}) = \left(\begin{array}{c c c c c}
          0 & 0 & 0 & 0 & 0  \\
          0 & 0 & 0 & 0 & 0  \\
          0 & 0 & 0 & 0 & 0  \\
          0 & 0 & 0 & 0 & 0  \\
          \tfrac{5\mathcal{K} u}{\rho} + \tfrac{10pu}{\rho^2} - \tfrac{10qu^2_x}{\rho}& -\tfrac{5\mathcal{K} }{\rho} + \tfrac{20qu}{\rho} & -\tfrac{10q}{\rho} & 0 & 0  \\
    \end{array}\right).
\end{equation}
For the numerical method, instead, we use the non-conservative product in primitive variables, where the matrix reads
\begin{equation}
    \hat{\mathbf{B}}(\mathbf{P}) = \mathbf{B}(\mathbf{P})\frac{\partial \mathbf{M}}{\partial \mathbf{P}} = \left(\begin{array}{c c c c c}
          0 & 0 & 0 & 0 & 0  \\
          0 & 0 & 0 & 0 & 0  \\
          0 & 0 & 0 & 0 & 0  \\
          0 & 0 & 0 & 0 & 0  \\
          \tfrac{10pq}{\rho^2} & -5\mathcal{K} & -\tfrac{10q}{\rho} & 0 & 0  \\
    \end{array}\right).
\end{equation}
\subsection{Numerical fluxes in fluctuation form}
For the vectors an matrices of Eq.~\eqref{eq:spaceTimeDiscretizationRegGrad}, we use the Rusanov-type scheme for the approximate Riemann solver, as follows,
\begin{multline}\label{eq:Dpm}
        \mathbf{D}^\pm(\mathbf{M}^n_{L},\mathbf{M}^n_{R}) = \tfrac{1}{2}\left[\mathbf{F}^n_R - \mathbf{F}^n_L + \tilde{\mathbf{B}}(\mathbf{P}^n_L, \mathbf{P}^n_\star)(\mathbf{P}^n_\star - \mathbf{P}^n_L)+\tilde{\mathbf{B}}(\mathbf{P}^n_\star, \mathbf{P}^n_R)(\mathbf{P}^n_R - \mathbf{P}^n_\star) \right] \\
        \pm \tfrac{|\lambda^{max}|}{2}\left(\mathbf{M}^n_{R} - \mathbf{M}^n_{L}\right)
\end{multline}
where the primitive variables $\mathbf{P}$ and the Roe-type matrix $\tilde{\mathbf{B}}$ are given in Appendix \ref{sec:Notation1}. The intermediate state (with the Rusanov scheme) is computed as follows,
\begin{equation}\label{eq:Ustar}
    \mathbf{M}^n_\star = \tfrac{1}{2|\lambda^{max}|}\left[|\lambda^{max}|\left(\mathbf{M}^n_L+\mathbf{M}^n_R \right)-(\mathbf{F}^n_R - \mathbf{F}^n_L) -  \tilde{\mathbf{B}}(\mathbf{M}^n_{L},\mathbf{M}^n_{R})(\mathbf{M}^n_{R}- \mathbf{M}^n_{L}) \right],
\end{equation}
and the averaged value of the primitive variables $\mathbf{P}^n_\star$ are computed from the conservative variables $\mathbf{M}^n_\star$. Finally, the fluxes at the interface are computed with the reconstructed variables, as
\begin{equation}
   \mathbf{F}^-_{i+1/2} = \mathbf{F}(\mathbf{M}^n_L)~~~\text{and}~~~\mathbf{F}^+_{i-1/2} = \mathbf{F}(\mathbf{M}^n_R).
\end{equation}
The slopes of the extrapolated values of the primitive variables are computed as
\begin{equation}
\Delta\mathbf{P}^n_i = \text{minmod}\left(\mathbf{P}^n_{i+1}-\mathbf{P}^n_i,\, \mathbf{P}^n_i- \mathbf{P}^n_{i-1}\right).
\end{equation}
The left and right states at the $i+1/2$ wall interface are computed with this slope, as in the MUSCL scheme, i.e.,
\begin{equation}\label{eq:reconstructedP}
    \mathbf{P}^n_L = \mathbf{P}^n_i + \tfrac{1}{2}\Delta\mathbf{P}^n_i~~~\text{and}~~~\mathbf{P}^n_R = \mathbf{P}^n_{i+1} - \tfrac{1}{2}\Delta\mathbf{P}^n_{i+1}.
\end{equation}
The conservative variables at the cell interface $\mathbf{M}^n_{L/R}$ are computed from the reconstructed primitive variables of Eq.~\eqref{eq:reconstructedP}. 
The Roe-type matrix of the non-conservative product is easily obtained by doing the arithmetic mean between the left and right states, as follows:
\begin{equation}
    \tilde{\mathbf{B}}(\mathbf{P}_R,\,\mathbf{P}_L) = \hat{\mathbf{B}}(\bar{\mathbf{P}})~~~\text{with}~~~\bar{\mathbf{P}} = \tfrac{1}{2}\left(\mathbf{P}_R + \mathbf{P}_L\right).
\end{equation}

\bibliography{references}

\end{document}